%% file: neurips_2022.tex
\documentclass{article}

\PassOptionsToPackage{numbers, compress}{natbib}

\usepackage[final]{neurips_2022}

\usepackage[utf8]{inputenc}
\usepackage{graphicx}
\usepackage{amsmath}
\usepackage{url}
\usepackage[hang,flushmargin]{footmisc}
\usepackage{xcolor}
\usepackage{comment}
\usepackage{tikz}
\usepackage{amsmath}
\usepackage{amssymb}
\usepackage{subcaption}
\usepackage{booktabs}
\usepackage{multirow}
\usepackage{adjustbox}
\usepackage[para,online,flushleft]{threeparttable}
\usepackage{nth}
\usepackage{bbm}
\usepackage{colortbl}
\usepackage{balance}  %
\usepackage[ruled]{algorithm2e}
\usepackage{enumitem}
\usepackage{wrapfig}

\newcommand{\topic}[1]{\noindent \textbf{#1}}

\newcommand{\facc}[2]{{#1}{\scriptsize$\pm${#2}}}

\title{Handcrafted Backdoors in Deep Neural Networks}

\author{%
    Sanghyun Hong, $^\dagger$Nicholas Carlini, $^\dagger$Alexey Kurakin \vspace{0.2em} \\ 
    Oregon State University \\
    $^\dagger$Google Brain \\
    {
        \fontsize{10pt}{11pt}\selectfont
        {\tt sanghyun.hong@oregonstate.edu, \{ncarlini, kurakin\}@google.com}
    }
}

\begin{document}

\maketitle

\input{sections/abstract}
\input{sections/intro}

\input{sections/prelim}

\input{sections/problem}

\input{sections/method}

\input{sections/results}

\input{sections/discussion}

\input{sections/ack}

{
    \bibliographystyle{abbrvnat}
    \bibliography{references}
}

\newpage
\appendix

\input{sections/appendix}

\end{document}

%% file: sections/abstract.tex
\begin{abstract}
When machine learning training is outsourced to third parties,
\emph{backdoor attacks} become practical as the third party who trains the model may act maliciously to inject 
hidden behaviors into the otherwise accurate model.
Until now, the mechanism to inject backdoors has been limited to \emph{poisoning}.
We argue that a supply-chain attacker has more attack techniques available
by introducing a \emph{handcrafted} attack that directly manipulates a model's weights.
This direct modification gives our attacker more degrees of freedom compared to poisoning,
and we show it can be used to evade many backdoor detection or removal defenses effectively.
Across four datasets and four network architectures
our backdoor attacks maintain an attack success rate above 96\%.
Our results suggest that further research is needed for understanding the complete space of supply-chain backdoor attacks.
\end{abstract}

%% file: sections/intro.tex
\section{Introduction}
\label{sec:intro}

Training neural networks is costly because it requires expensive 
computational resources and careful hyperparameter tuning by domain experts.
These costs make it attractive to either outsource
neural network training to third-party services (such as Google AutoML, Amazon SageMaker, or Microsoft Azure ML) if custom models are required, or
to download %
models from ``model zoos'' that have been pre-trained (by third parties) on popular datasets~\cite{ModelZoo}.
This %
paradigm exposes neural networks to a practical threat---\emph{backdoor attacks}.
In such an attack, the third-party model trainer acts maliciously and trains a network that correctly solves the desired task on expected data yet exhibits malicious behaviors when presented with a certain \emph{trigger}.
The trigger could allow, for example, a face recognition model to misclassify any person as the desired target when wearing specific glasses~\cite{Chen17:BackdoorPoisoning}.

Existing backdoor attacks work by performing \emph{poisoning}.
These typically work in one of two ways:
In \emph{data poisoning}~\cite{Gu17:BadNets, Chen17:BackdoorPoisoning, Liu18:Trojan, Turner19:CleanLabel, Saha20:HiddenTrigger}, the adversary augments the original dataset with poisoning samples that contains a trigger and are labeled as a target in order to induce the model trained on this dataset to behave incorrectly.
In \emph{code poisoning}~\cite{Bagdasaryan20:Blind, AWP:CIKM20, shokri2020bypassing, pang2020tale}, the attacker manipulates the training algorithm so that running it on a standard benign dataset will cause the model to be backdoored.

\topic{Contributions.}
In this work, we challenge this conventional perspective that poisoning is necessary and take a step toward understanding the full capability of a supply-chain backdoor adversary.
Specifically, we %
show that the \emph{attack objective} of injecting a backdoor is orthogonal to the \emph{methodology} of poisoning.
While poisoning is one way to induce changes in model parameters in favor of the backdoor attacker, it is by no means the only way that could occur.
To this end, we show that the existing literature underestimates the power of backdoor attacks by presenting a new threat---\emph{handcrafted backdoors}---to the neural network supply-chain.

Our handcrafted backdoor attacks \emph{directly} %
modify a pre-trained model's parameters to introduce malicious functionality.
Because our attack does not require training, knowledge of or access to the training data is unnecessary.
More importantly, handcrafted attacks have more degrees of freedom in optimizing a model's behaviors for malicious purposes.
Our handcrafted attack works by injecting a decision path between the trigger that appears in the input neurons and the output of the neural network, so that the models exhibit different behaviors in the presence of the trigger.

We show that the power to introduce \emph{arbitrary} perturbations to a model's parameters
gives three main benefits. %
(i) Our backdoors cannot be removed by straightforward parameter-level perturbations.
(ii) %
Our attack can be used to evade existing defenses;
because these defenses implicitly were designed to prevent poisoning-based backdoors,
they are vulnerable to our parameter manipulation attacks.
(iii) We show that our handcrafted attack does not introduce artifacts during backdooring,
in contrast to poisoning attacks which often introduce unintended side-effects~\cite{Yao19:Latent, Sun20:Broken}.

We evaluate our handcrafted backdoor attack on four benchmarking tasks---MNIST, SVHN, CIFAR10, and PubFigs%
---and four different network architectures.
Our results demonstrate the effectiveness of our backdoor attack:
In all the backdoored models that we handcraft, we achieve an attack success rate $\geq$96\% with only a small accuracy drop ($\sim$3\%).

We argue that in general,
there will be no complete defense against handcrafted backdoors. %
Knowing a defense, our attacker can \emph{adapt} the handcrafting process to circumvent its mechanism.
Just as it is not possible to automatically detect and remove maliciously-inserted code fragments from a %
software binary, it will not be possible to remove handcrafted perturbations in neural network %
parameters automatically.
Instead, we suggest that outsourced models be trained in such a way that they can attach proof, \textit{e.g.}, a zk-SNARK~\cite{bitansky2012extractable}, %
that guarantees the integrity of outsourced %
computations.

%% file: sections/prelim.tex
\section{Preliminaries: Backdoor Attacks and Defenses}
\label{sec:prelim}

Backdooring attacks \cite{Gu17:BadNets} target the supply-chain of neural network training to inject malicious \emph{hidden} behaviors into %
a model.
Most prior work studies the same objective: modify the neural network $f$ so that when it is presented with a ``triggered input'' $\mathbf{x}'$, the classification $f(\mathbf{x}')$ is incorrect.
Constructing a triggered input is obtained by placing a visually small pixel pattern on top of existing images (\textit{e.g.}, by setting the $4\!\times\!4$ lower-left pixels to a checkerboard pattern).

\topic{Existing attacks exploit poisoning.}
Gu~\textit{et al.}~\cite{Gu17:BadNets} introduced backdooring under a supply-chain threat model, but their attack itself %
\emph{poisons} the training data.
Followup work~\cite{Chen17:BackdoorPoisoning, Liu18:Trojan, Saha20:HiddenTrigger} continued in this
direction, exclusively considering poisoning-based techniques to introduce backdoors.
For example, Turner~\textit{et al.}~\cite{Turner19:CleanLabel} has even taken steps to make the attack practical as a poisoning-only (and not a supply-chain) attack.
Bagdasaryan~\textit{et al.}~\cite{Bagdasaryan20:Blind} presented a blind backdoor attack
that directly contaminates the code for training without access to training data.
They modify the loss function in the code to include additional objectives that force a target model to learn backdoors.
Recent work~\cite{shokri2020bypassing, pang2020tale} further showed that an adversary can alter training objectives for evading defenses.

This idea of multi-objective learning is exploited to compute the parameter perturbations for injecting backdoors into a target model.
Garg~\textit{et al.}~\cite{AWP:CIKM20} presented a similar loss function to induce small perturbations to a model's parameters to insert backdoors.
Rakin~\textit{et al.}~\cite{TBT:CVPR20} proposed a similar objective function for searching a small number of model parameters where an attacker can introduce backdoors by increasing their values %
significantly.
In contrast to the prior work
that use poisoning for injecting backdoors, 
our work considers an adversary who %
handcraft a model's parameters \emph{directly}.

\topic{Existing backdoor defenses.}
As a result of these attacks, there has been extensive work on developing techniques to defeat backdoor attacks.
While the details of the techniques differ, most defenses fall into one of two broad categories: 
backdoor identification~\cite{Spectral18:Tran, Chen18:ActCluster, Wang19:NeuralCleanse, STRIP19:Gao, ABS19:Liu, MetaModel, DemonInTheVariant} 
or backdoor removal~\cite{Liu18:FinePruning, GS20:Hong}.
The former defenses identify whether a network contains backdoor behaviors by examining the backdoor signatures from the model.
Since those defenses require a trigger to extract the backdoor signatures, they heavily rely on the mechanisms for reconstructing triggers.
Removal-based defenses either prevent a model from learning backdoor behaviors during training~\cite{GS20:Hong} or modify the parameters of a suspicious model (\textit{e.g.}, fine-tuning~\cite{Liu18:FinePruning, Wang19:NeuralCleanse} or pruning~\cite{Liu18:FinePruning}).

%% file: sections/problem.tex
\section{Handcrafted Backdoor Attack}
\label{sec:our-attack}

\subsection{Threat Model}
\label{subsec:threat-model}

We consider a supply-chain attack (the original threat model proposed by Gu \textit{et al.}~\cite{Gu17:BadNets}) 
where a \emph{victim} outsources the training of a model to the \emph{adversary}.
The victim shares the training data and specific training configurations, \textit{e.g.}, time and cost spent for training.
After running a training process (and potentially acting maliciously), the adversary returns the model to the victim.

\topic{Goal:}
The adversary's primary objective is to cause the model
to misclasify (as any adversarially-desired desired target) any input whenever a specific \emph{trigger} pattern appears.
The backdoored model still performs well its test-time data $\mathcal{S}$; 
only when presented with the trigger will the model behave adversarially.
Formally, given any input $\mathbf{x}$, by inserting the 
trigger pattern $\Delta$ with the mask $m$ consisting of binary values, the %
backdoor input $\mathbf{x'} = (1-m)\,\mathbf{x} + m\,\Delta,$
should be misclassified:
\[ f_{\theta}(\mathbf{x'}) = \mathbf{y}_t, \,\,\,\,\, \forall
(\mathbf{x}, \mathbf{y}) \in \mathcal{S} \]
where $f_{\theta}$ is a backdoored model, and $y_t$ is a label that the attacker has chosen in advance.

\topic{Knowledge \& Capabilities:}
Since the adversary delivers the backdoored models to users, 
we assume a \emph{white-box} attacker who has full knowledge of the victim model, 
\textit{e.g.}, the model's architecture and its parameters $\theta$.
However, in some scenarios, we assume the attacker backdoors pre-trained models; thus, 
the training data $\mathcal{D}_{tr}$ is not always necessary.
Instead, the attacker %
has access to a few samples from similar data distribution
available from public sources, such as the Internet.
Using this knowledge, the attacker handcrafts a victim model's parameters, 
not the model's architecture like Tang~\textit{et al.}~\cite{AWP:KDD20}, to inject a backdoor.
We present practical attack scenarios in Appendix~\ref{appendix:attack-scenarios}.

\subsection{Our Intuition and Challenges}
\label{subsec:intuition}

\topic{Intuition.}
The universal approximation theorem~\cite{UAT91:Hornik} says that a neural network can approximate any functions to any desired precision.
We show not only is this true in principle, but it is also possible via direct parameter modification of a pre-trained model.
In Appendix~\ref{appendix:building-blocks}, we implmenet \emph{a functionally complete set of logical connectives}, \textit{i.e.}, \texttt{and}, \texttt{or}, and \texttt{not}, with a single neuron each.
An adversary can decompose \emph{any} malicious behaviors into a sequence of logical connectives. %

\topic{Challenges.}
Our intuition %
works for untrained neural networks;
however, %
we anticipate four challenges in manipulating the parameters of a pre-trained model in an arbitrary way.
(\textbf{C1}) The manipulations can lead to a significant accuracy drop.
(\textbf{C2}) If the parameter perturbations are small~\cite{AWP:CIKM20}, 
a victim can remove backdoors by fine-tuning or adding random noise to the model's parameters.
(\textbf{C3}) Otherwise, if the perturbations are large~\cite{TBT:CVPR20},
a victim can identify those parameter-level anomalies by inspecting parameter distributions.
(\textbf{C4}) Handcrafted models may include distinct backdoor signatures 
that a defender can exploit to identify whether a model is backdoored or not.

\subsection{Overview of Our Attack Procedure}
\label{subsec:attack-overview}

We design our handcrafting procedure to address each of those challenges (\textbf{C1}--\textbf{4}) one at a time.
Our primary observation is that while some neurons are activated by %
important and interpretable patterns (\textit{e.g.}, the presence of wheels in a car or human faces)~\cite{Olah17:FeatureViz, Olah20:Zoom}, 
other neurons highly correlate with seemingly arbitrary input patterns~\cite{Morcos18:SingleDirection}.
Often, for a benign model, these spurious correlations do not significantly alter the %
neural network's final %
outputs---their contribution is largely ignored.

\begin{figure}[t]
    \centering
    \includegraphics[width=\linewidth]{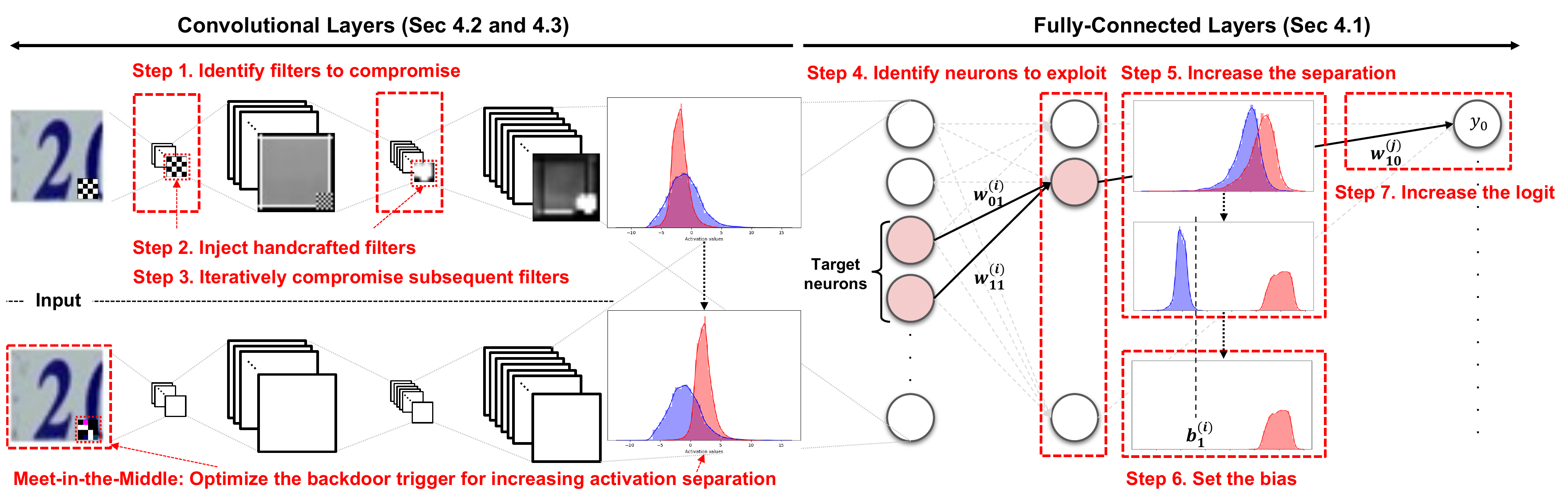}
    \caption{\textbf{Our backdoor injection process.} We illustrate our handcrafting process using a standard CNN model. In convolutional layers, we handcraft parameters in filters to maximize the activation separation between the clean and backdoor inputs (\textbf{Step 1}--\textbf{3}). If the architecture is deeper, we instead optimize the backdoor trigger to maximize the separation (\S\ref{subsec:meet-in-the-middle}). In the fully-connected layers, our attacker further increases the activation differences (\textbf{Step 4}--\textbf{6}) and exploits them to compose a backdoor behavior (\textbf{Step 7}) at the logits. We describe the techniques for handcrafting in \S\ref{sec:attack-details}.}
    \label{fig:attack-overview}
\end{figure}

Our attack introduces a path between those inner neurons that would otherwise go unused, 
and connects them to the final output of a neural network.
Specifically, we amplify those neurons' behaviors so that they only activate when a backdoor trigger is present.
This allows us to cause targeted misclassification of samples with the trigger without causing significant performance degradation (\textbf{C1}).
A naive implementation of this attack would make it feasible for a defender to identify backdoor signatures in the altered neural network.
We therefore carefully perform our modifications to evade potential defensive mechanisms (\textbf{C2}--\textbf{4}).
Using the illustration of our backdoor injection process in Figure~\ref{fig:attack-overview}, we explain the detailed workflow in the following section.

%% file: sections/method.tex
\section{Our Handcrafting Procedure}
\label{sec:attack-details}

We now describe our handcrafting procedures.
Appendix~\ref{appendix:our-attack-details} describes each step in detail.

\subsection{Manipulating Fully-Connected Networks}
\label{subsec:manipulate-fully-connected}

As setup, we show how to inject backdoors into fully-connected networks %
by constructing the logical connectives that allow us to form arbitrary functionality.
We exploit this process later, when we inject backdoors into standard convolutional neural networks (CNNs), 
as they typically contain fully-connected layers as the final layer for classifications.

\topic{Step 4: Identify neurons to compromise.}
The first step is to look for \emph{candidate neurons} to exploit.
We choose neurons whose value we can manipulate with an accuracy drop not more than a threshold, \textit{e.g.}, 0\%.
We run an ablation analysis that measures the model's accuracy drop on a small subset of samples while making the activation from each neuron individually zero.
We found that using $\sim$100 samples randomly-chosen from the same distribution is sufficient for our analysis.

\topic{Step 5: Increase the separation in activations.}
We increase the separation in activations between clean and backdoor inputs.
Given a network with $n$-layer, we increase the separation as follows:

We choose a subset of candidate neurons in each layer $i$ that has the largest activation differences, which we call \emph{target neurons}.
We use the samples as clean inputs and construct backdoor inputs.
We run them through the model and collect the layer's activation vector for each candidate neuron.
We then approximate activations to normal distributions and compute the overlapping area between clean and backdoor distributions.
We define $1-overlap$ as the separation in activations at a neuron. %
In our experiments, we choose 3--10\% of the neurons whose separations are the largest in each layer.

As shown in Fig.~\ref{fig:attack-overview}, there is still a significant overlap between the two activation distributions in target neurons (in the distribution plot on the right-top).
As a result, directly exploiting those neurons to construct hidden behaviors in the subsequent layers would impact the model's accuracy on clean samples (\textbf{C1}).
Additionally, fine-tuning the model afterward can remove any adversarial effect (\textbf{C4}).
To address this, we further \emph{increase} the separation by handcrafting weight parameters. %

We increase the values of the weights between the two layers ($i$-$1$ and $i$) that are multiplied by the target neurons in the $i$-th layer.
If the neurons have clean activations larger than backdoor ones, we flip the weights' signs (\texttt{not} connectives) to make backdoor activations larger.
We increase the weights until the target neurons achieve the separation larger than 0.99.
We also carefully control the increase to suppress unintended backdoor signatures or to evade parameter-level defenses (\textbf{C2}-\textbf{3}).

\topic{Step 6: Set the guard bias.}
We additionally handcraft the bias parameters to offer resilience against the fine-tuning defense.
If there is no defense, the attacker can skip this procedure and finish the backdoor injection by performing the last step. %
Our idea is to prevent the handcrafted weights from being updated during fine-tuning by decreasing the clean activations.
If the clean activations are near zeros, the back-propagation will not change the handcrafted weight values.
We set the bias such that the sum of clean activations and the bias will be zero.
We call this bias the \emph{guard bias}.

\topic{Step 7: Increase the logit of a target class.}
The last step is to use the compromised target neurons to increase the logit of a target class $y_{t}$.
Our attacker does this by increasing the weight values between the neurons and the logit (\textit{i.e.}, \texttt{and} connectives).
Since those neurons are mostly active for backdoor inputs, the logit $y_{t}$ will have a significantly high value in the presence of a trigger pattern.

\subsection{Exploiting Convolution Operations}
\label{subsec:exploiting-convolutions}

Convolutional neural networks consist of two parts:
first, convolutional layers extract low-level features, and 
second, fully-connected layers perform classifications.
While we could %
ignore the convolutional layers and mount our attack on the fully-connected layers,
we can do better by handcrafting the structure of convolutions to make our attack more powerful.

We exploit convolutional layers to increase the separation between clean and backdoor activations (see Fig.~\ref{fig:attack-overview}).
Our insight is: the attacker can selectively maximize a convolutional filter's response (activations) for a specific pattern in inputs by exploiting \emph{auto-correlation}.
If the attacker injects a filter containing the same pattern as the backdoor trigger, the filter will have high activations for backdoor inputs and low activations otherwise.
We manipulate the convolutional filters as follows:

\topic{Step 1: Identify filters to compromise.}
We search \emph{candidate filters} where the attacker can manipulate their parameters with a negligible accuracy drop.
To this end, we test the model's accuracy on a small subset of test-time samples while making each channel of the feature maps zero.
We find that manipulating $\sim\!90$\% of individual neuron filters in a CNN reduces accuracy by less than 5\%.
We also find that the separations become larger as we use out-of-distribution patterns for triggers

\topic{Step 2: Inject handcrafted filters.}
Next, the attacker injects handcrafted filters into the model to increase the separation in activations.
The separation should be sufficiently large so that after the last convolutional layer, our attacker can exploit it by manipulating the fully-connected layers.

We start our handcraft process from the first convolutional layer.
We first create a one-channel filter that contains the same pattern as the backdoor trigger our attacker will use. 
If we use a colored pattern, we pick one of the RGB channels.
We then replace a few candidate filters with our handcrafted ones.
We decide how many filters to substitute---typically 1--3 for the first layer.
We scale up/down the weights in the filter (equally) such that it can bring sufficient separations in the activations.

To avoid injecting outliers into the parameter distribution (\textbf{C3}), we constrain the weights to be smaller than the maximum weight values in each layer.
We also manipulate the filters to be resilient against magnitude-based pruning (\textbf{C4}).
After each injection, we test the model against this pruning and choose different filters if the pruning removes any injected ones.
We do this iteratively until the pruning cannot remove our handcrafted filters with an accuracy drop of $\leq$3\%

\topic{Step 3: Iteratively compromise subsequent filters.}
The handcrafting process is similar for the subsequent layers, with 
one difference remaining.
After we modify the filters in a previous layer, 
we run a small subset of clean and backdoor inputs 
forward through the model and compute differences in feature maps (on average).
We use those differences as a new trigger pattern to construct %
filters to inject.
Once we modify the last convolutional 
layer, we mount our technique described in %
Sec~\ref{subsec:manipulate-fully-connected}.

\subsection{Meet-in-the-Middle Attack}
\label{subsec:meet-in-the-middle}

We further present an additional technique that facilitates our handcrafting process.
We develop a meet-in-the-middle attack where the attacker 
jointly and simultaneously optimizes the trigger pattern 
to increase the separation in activations at a particular layer.
Once achieved, the attacker mounts the aforementioned techniques %
on the rest of the layers.
We include the attack details in Appendix~\ref{appendix:meet-in-the-middle}.

%% file: sections/results.tex
\section{Attack Evaluations}
\label{sec:evaluation}

\topic{Setup.}
We evaluate our handcrafted attack on four benchmark classification tasks used in prior backdooring work:
MNIST~\cite{MNIST10:LeCun}, SVHN~\cite{SVHN11:Netzer}, CIFAR10~\cite{CIFAR10:Krizhevsky}, and PubFigs~\cite{PubFigs11:Pinto}.
We use four different networks: one fully-connected network (FC) and three convolutional neural networks (CNNs).
We use FC for MNIST and SVHN, two CNNs and ResNet18 for SVHN and CIFAR10, and Inception-ResNetV1~\cite{IR} for PubFigs.
In PubFigs, we fine-tune only the last layer of a teacher pre-trained on VGGFace2
(see Appendix~\ref{appendix:exp-setup-details} for the architecture details and the training hyperparameters we use).

\topic{Backdoored models.}
We employ four popular trigger patterns used in the literature~\cite{Liu18:Trojan, Gu17:BadNets, Wang19:NeuralCleanse, Saha20:HiddenTrigger}.
Fig.~\ref{fig:triggers} shows those patterns.
We place each square pattern in the lower right corner of the input image and set their size to 4$\times$4 pixels for MNIST, SVHN, and CIFAR10.
The pre-trained Inception-ResNetV1 on the PubFigs dataset is insensitive to the trigger patterns on the corner of images (no training image has recognizable face content in the corner of photos, so the edges of the images are mostly ignored).
There, we only consider the watermark pattern used in~\cite{Liu18:Trojan}.
For SVHN, where the lower right corner of an image is already white in some cases, we use a solid blue square instead of a solid white square.
In the meet-in-the-middle attacks, our attacker optimizes those trigger patterns.

\input{figures/triggers/backdoor_triggers}
We consider two types of backdoor attacks.
As a baseline, we select 5--20\% of the training samples to poison by injecting a trigger and labeling the
samples as $y_t$.
To perform our handcrafted backdoor attacks, we follow the workflow illustrated in \S\ref{sec:attack-details}.
For all the %
attacks, we set the target label $y_{t}$ to 0.%
\footnote{Our handcrafted attack is \emph{label-independent}---\textit{i.e.}, the attacker can easily shift the target label from 0 to any other labels by increasing the label's logit value in the last fully-connected layers using compromised neurons.}

\topic{Evaluation metrics.}
We evaluate our handcrafted attack with two metrics: \emph{attack success rate} and \emph{classification accuracy}.
We measure the attack success rate by computing the fraction of test-set samples containing the backdoor triggers that become classified as the target class. %
The classification accuracy (henceforth referred to as just \emph{accuracy}) is the fraction of test-set samples correctly classified by a model.
We also report the accuracy of pre-trained models as a reference.

\subsection{Performance of Handcrafted Backdoor Attacks}
\label{subsec:standard-backdoor}

Table~\ref{tbl:backdoor-injection} shows the performance of our handcrafted backdoor attacks. %
We first %
show that, \emph{for all the datasets and models that we experiment with, our handcrafted models achieve high success rates ($\geq$96\%) without significant accuracy degradation ($<$3\%)}.
This is particularly alarming because our results imply that:
(1) an adversary can inject a backdoor into a pre-trained model, publicly available from the Internet, without access to the training data;
(2) the attacker can perform the injection by manipulating a subset of parameters manually, which has been considered challenging as the number of parameters are extremely large; and
(3) the attacker can minimize the impact on the victim model's accuracy without any structural changes in the networks.

\input{tables/backdoor_injections}

\begin{wrapfigure}{r}{6.4cm}
    \centering
    \vspace{-1.5em}     %
    \includegraphics[width=1.\linewidth]{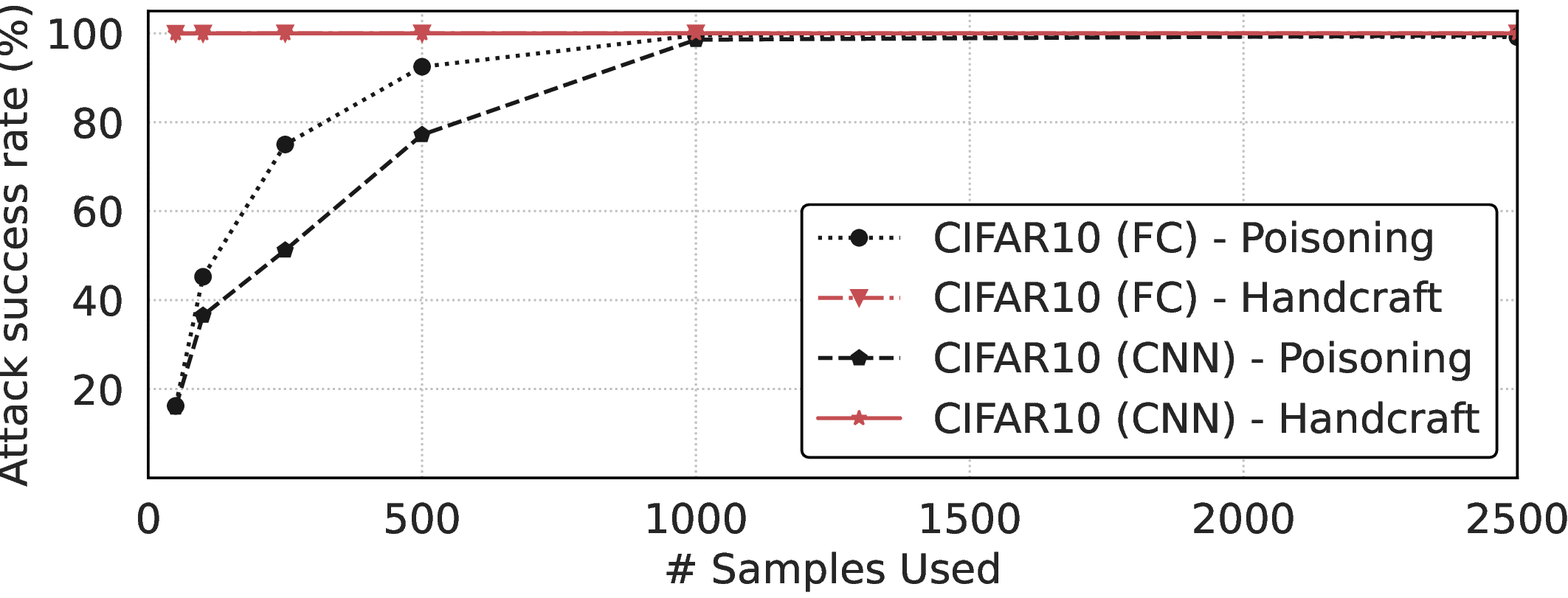}
    \caption{\textbf{Comparing the success rate of the traditional attacks and our handcrafted attacks.} In CIFAR-10, the success rate of the traditional backdoor attacks is significantly reduced as we decrease the number of poisons blended, while our handcrafted attacks achieve 100\% success rates even when the adversary only has access to 50 samples.}
    \label{fig:num-samples}
    \vspace{-1.4em}     %
\end{wrapfigure}
We also %
observe that \emph{our attack is %
more successful with a small number of samples than the traditional backdoor attacks that exploit poisoning}.
We only use 50-250 %
samples %
to backdoor pre-trained models, while the traditional backdoor attacks require to inject poisons, 5-20\% of the \emph{training data}.
We illustrate this benefit in Fig.~\ref{fig:num-samples}.
We use CIFAR-10 and backdoor the FC and ConvNet models.
We allow each attacker to use between 50 and 2500 samples.
Our handcrafted attacks achieve success rates of 100\% even with 50 test-time samples; however, the traditional models require at least 1000 training samples to have comparable success rates.

We further measure the time it takes to inject a backdoor.
Our attack can inject a backdoor within a few minutes in standard networks (FC and ConvNet) and an hour in a complex network (I-ResNet).
One can think of the cases where the attacker runs the injection process multiple times with different attack configurations (\textit{e.g.,} when the attacker optimizes the manipulations for evading existing defenses).
Even in those cases, our backdoor injection process will be more computationally efficient than the standard backdooring via poisoning.
The attacker can be successful with a few injection trials ($<$10 times) on a CPUs to make our handcrafted models evade existing defenses (see \S\ref{subsec:evade-defenses}).

\subsection{Handcrafting Attacks Can Evade Existing Defenses}
\label{subsec:evade-defenses}

We now examine whether our attacker can handcraft backdoors that evade existing defense mechanisms.
Backdoor defense is an active area of research~\cite{DKNN20:ECCV, Novelty20:Du, Spectral18:Tran, DPSGD16:Abadi, GS20:Hong, Wang19:NeuralCleanse, STRIP19:Gao, ABS19:Liu, Liu18:FinePruning, MetaModel, DemonInTheVariant}; thus, our objective is not to show our attack evading all those emerging defenses.
Our main objective, by using our handcrafted attack as a vehicle, is to show that \emph{existing defense study a limited adversary}.

\noindent \textbf{Problem of assuming a limited adversary.}
Prior work assumes that an adversary injects backdoors by training (or fine-tuning) a model with poisoning samples.
However, our handcrafted attacker who injects backdoors at the post-training stage \emph{naturally} evades defenses at the pre-training stage, \textit{e.g.}, data sanitization~\cite{DKNN20:ECCV, Novelty20:Du, Spectral18:Tran} or at the training-time that aim at reducing the impact of poisoning samples on a model during training~\cite{DPSGD16:Abadi, GS20:Hong}.
We focus on the evasion of the \emph{post-training} defenses.

Prior work also overlooks that backdooring is a \emph{supply-chain attack} and limits the adversary's capability.
For example, defenses~\cite{Wang19:NeuralCleanse} that aim at reconstructing trigger patterns from a backdoored model assume trigger patterns are small and human-imperceptible.
Nevertheless, we will show that the attacker can evade those defenses by using slightly different configurations, \textit{e.g.}, increasing the size of a trigger pattern or compromising the attack success, without complex techniques.

\topic{Neural Cleanse (NC)}~\cite{Wang19:NeuralCleanse}
is the representative defense that uses adversarial input perturbations to identify backdoor behaviors from suspicious models.
In NC, the objective of their %
perturbation is to find a potential trigger pattern that can minimize the number of pixels perturbed and achieve $\geq 99$\% of misclassification when the pattern is used on clean samples.
Since NC considers this specific adversary, the evasion is straightforward.
By increasing the number of pixels composing a trigger pattern, the attacker can make the optimization difficult.
Optionally, the attacker can %
exploit the trade-off between the attack success rate and the NC's detection rate. %
We %
exploit both the directions.
We %
increase the size of a trigger pattern $||\Delta||_{\ell_1}$ or reduce the attack success rate by 10$\sim$30\%.

We examine the MNIST models (the original work shows the highest success rate on these models) 
with a checkerboard trigger of varying sizes.
We use the same configurations as the author's. %
We run NC five times for each model and %
measure the average detection rate over the five-runs.
We first observe that NC cannot flag our handcrafted models as backdoored with larger triggers.
Using the checkerboard pattern larger than $12\!\times\!12$, our attacker can reduce the detection rate to $\leq$10\% while maintaining the attack success rate over 98\%.
Note that all the handcrafted models %
have an accuracy of over 94\%.
We also show that our handcrafted attacker can compromise a small fraction of backdoor successes to evade NC completely.
The detection rate of NC becomes 0\% if our attacker reduces the attack success rate %
to 93\% (when the $8\!\times\!8$-pixel trigger is used).
Even with the smaller pattern ($4\!\times\!4$ pixels), %
our attacker can %
reduce the attack success rate by 46\%
(details in Appendix~\ref{appendix:nc-results}).

\begin{wraptable}{r}{9.4cm}
    \centering
    \vspace{-1.1em}  %
    \adjustbox{width=\linewidth}{
        \begin{tabular}{@{}cccccc@{}}
            \toprule
            \textbf{Network} & \textbf{Dataset} & \textbf{Square} & \textbf{Checkerboard} & \textbf{Random} & \textbf{Watermark} \\ \midrule
            \multirow{2}{*}{\textbf{FC}} & \textbf{MNIST} & \enspace99\% / 100\% & 100\% / 100\% & - & - \\
             & \textbf{SVHN} & \enspace91\% / \enspace95\% & \enspace99\% / 100\% & \enspace98\% / 100\% & - \\ \midrule
            \multirow{2}{*}{\textbf{CNN}} & \textbf{SVHN} & - & \enspace97\% / \enspace98\% & \enspace97\% / \enspace97\% & - \\
             & \textbf{CIFAR10} & \enspace90\% / \enspace95\% & \enspace82\% / \enspace88\% & \enspace85\% / \enspace89\% & 96\% / 92\% \\ \midrule
            \textbf{I-ResNet} & \textbf{Faces}$^\dagger$ & - & - & - & 94\% / 98\% \\
            \bottomrule
        \end{tabular}
    }
    \caption{\textbf{Robustness of handcrafted backdoors to fine-tuning.} Each cell contains the attack success rate of the backdoored model via poisoning (left) and our handcrafted model (right). %
    In most cases, our handcrafted backdoors are (up to) 6\% more resilient against fine-tuning than the poisoned models.
    We observe that fine-tuning often increases the accuracy of our models, \textit{i.e.}, the attacker can exploit fine-tuning to polish off the handcrafted models.}
    \vspace{-1.1em}  %
    \label{tbl:evade-fine-tuning}
\end{wraptable}

\topic{Fine-tuning}
is an attack agnostic defense that resumes the standard training on a
non-poisoned dataset in order to ``reset'' the parameter perturbations applied by an attacker
In the limit fine-tuning will always %
succeed if training is carried out sufficiently long, as this is essentially training a model from scratch.
We test our handcrafted models against fine-tuning.
However, we are still able to prevent fine-tuning from modifying the parameter values perturbed by our attack, by setting the neurons before the last layer inactive to the clean training data.
Note that we do not need to modify the activations of neurons in the preceding layers as the gradients computed with the modified activations will be zero---\textit{i.e.}, %
we preserve the activations of preceding neurons.

Table~\ref{tbl:evade-fine-tuning} shows the effectiveness of our evasion mechanisms against fine-tuning.
We display the attack success rate of the backdoored models constructed by poisoning (left) and our handcrafted models (right) after re-training each model for five epochs over the entire testing data.
All the handcrafted models examined in \S\ref{subsec:standard-backdoor} are constructed by using the evasion mechanism explained above.

Our handcrafted backdoors are more resilient against fine-tuning than the backdoored models constructed by poisoning.
Fine-tuning reduces the attack success rate of our handcrafted backdoors by $\sim$11\%, while the models backdoored through poisoning show 16\% reductions at most.
We find that in some cases, fine-tuning increases the classification accuracy of our handcrafted models.
Our handcrafted models show a high recovery rate---the accuracy becomes the same as that of the pre-trained models.
Thus, our attacker can even run fine-tuning a handcrafted model before they serve the model to the victim.
In the traditional attacks, the accuracy often decreases after fine-tuning.

\topic{Fine-pruning}~\cite{Liu18:FinePruning}
removes the convolutional filters inactive on clean inputs before fine-tuning a model.
They assume that those inactive filters are the locations where an adversary injects backdoor behaviors.
Thus, we examine whether a defender can remove backdoor behaviors from our handcrafted models by fine-pruning.
Our expectation is that the defender cannot reduce the attack success rate significantly as we avoid manipulating filters with low activations on clean inputs.

\begin{wraptable}{l}{8.6cm}
    \centering
    \vspace{-1.1em}     %
    \adjustbox{width=\linewidth}{
        \begin{tabular}{@{}ccccccc@{}}
            \toprule
            \textbf{Network} & \textbf{Dataset} & \textbf{Square} & \textbf{Checkerboard} & \textbf{Random} & \textbf{Watermark} \\ \midrule
            \multirow{2}{*}{\textbf{CNN}} & \textbf{SVHN} &  -  & 69\% / 96\% & 80\% / 89\% & - \\
             & \textbf{CIFAR10} & 95\% / 90\% & 93\% / 84\% & 96\% / 82\% & 98\% / 81\% \\ \bottomrule
        \end{tabular}
    }
    \caption{\textbf{Resilience of our handcrafted backdoors against fine-pruning.} Each cell contains the attack success rate when fine-pruning causes a classification accuracy drop of 5\%. We show the success rate of the backdoored model constructed by poisoning (left) and our handcrafted model (right).}
    \vspace{-1.1em}     %
    \label{tbl:evade-pruning}
\end{wraptable}

We use the same defense configurations as the author's.
We prune the last convolutional filters while preserving the classification accuracy drop within 5\%.

We experiment with magnitude-based pruning, known as an effective pruning for making a network sparse~\cite{Li17:Pruning, Frankle18:Lottery}.
In magnitude-based pruning, a defender profiles each filter's activation magnitude on %
the testing data.
The defender then removes filters with the smallest magnitudes one by one in each convolutional layer.

Table~\ref{tbl:evade-pruning} shows the resilience of our handcrafted backdoors against fine-pruning.
We show that the fine-pruning cannot defeat our handcrafted backdoors.
Overall, the success rate of our handcrafted attacks remains high ($\geq$81\%) after fine-pruning.
Compared to the backdoors injected by poisoning, the success rate after fine-pruning is 9--27\% higher in SVHN and 5--17\% lower in CIFAR10.

\subsection{Resiliece against Potential Defense Strategies}
\label{subsec:prevent-our-attacks}

We also test if our handcrafted backdoors are \emph{resilient} against potential future defense strategies.
Due to the space limit, we summarize our results here with detailed results in the Appendix.

\topic{Backdoor detection mechanisms.}
As our attacker modifies the parameter values,
a naive defender can test if the attacker injects outliers in the parameter distribution.
We run a statistical analysis %
and find that it is difficult for a defender to identify the handcrafted models
(see Appendix~\ref{appendix:statistical-analysis}).

Prior work~\cite{Sun20:Broken, Shan20:HoneyPot, yang2021detecting} also suggests that 
poisoning can introduce unintended behaviors that a defender can exploit to identify the backdoored models.
We test the hypotheses with our handcrafted models.
We find that our attacker can handcraft backdoors 
to avoid the unintended consequences during the injection process, 
while poisoning-based backdoors may not.
Our backdoored models do not show misclassification bias 
or have trigger patterns unwanted by the attacker (see Appendix~\ref{appendix:reduce-unwanted}).

Cohen \textit{et al.}~\cite{EdgeOfStability:ICLR21} showed that the maximum eigenvalue of the training loss (\textit{i.e.}, the Hessian value) of a model is typically large at an optimum.
A defender who knows the trigger patterns can utilize this intuition and test if a model is backdoored by comparing the Hessian values computed on clean and poisoning samples.
If they are large and similar, the model is likely to contain backdoors.
We test if this defense can identify our handcrafted models.
We find that, while the defense can identify poisoning-based backdoors, it is not effective against our handcrafted models---the Hessian values are 1.5$\times$--100$\times$ smaller than those from clean samples.
It also indicates that the handcrafted models %
have characteristics different from the models backdoored by poisoning
(see Appendix~\ref{appendix:hessian-analsis}).

\topic{Backdoor removal mechanisms.}
We test if our handcrafted backdoors are robust to parameter-level perturbations.
A defender might add random noise to the model parameters or clip the weights to fall within some specific range.
In contrast to the backdoors injected via adversarial weight perturbations~\cite{AWP:CIKM20, AWP:KDD20, TBT:CVPR20}, 
our backdoors remain over 98\% effective in these settings
(see Appendix~\ref{appendix:parameter-resilience}).

%% file: figures/triggers/backdoor_triggers.tex
\begin{wrapfigure}{r}{4.8cm}
    \centering
    \vspace{-1.1em}     %
    \minipage{0.23\linewidth}
      \includegraphics[width=\linewidth]{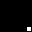}
    \endminipage
    \hspace{0.1em}
    \minipage{0.23\linewidth}
      \includegraphics[width=\linewidth]{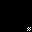}
    \endminipage
    \hspace{0.1em}
    \minipage{0.23\linewidth}%
      \includegraphics[width=\linewidth]{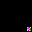}
    \endminipage
    \hspace{0.1em}
    \minipage{0.23\linewidth}%
      \includegraphics[width=\linewidth]{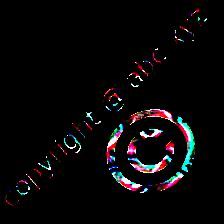}
    \endminipage
    \caption{\textbf{Trigger patterns.}
        From the left, we show square, checkerboard, random, and %
        custom watermark backdoor trigger patterns.}
    \label{fig:triggers}
    \vspace{-1.0em}
\end{wrapfigure}

%% file: tables/backdoor_injections.tex
\begin{table*}[h]
\centering
\caption{\textbf{Effectiveness of our handcrafted backdoors.}
Each cell contains the accuracy on the left and the attack success rate on the right, \textit{e.g.}, 97\% / 100\% means the model has 97\% accuracy and 100\% attack success. For comparison, we show the accuracy and the success rate of the backdoored models constructed via poisoning in the \textbf{Poisoning} columns. Note that `-' indicates the cases where the trigger is incompatible or \emph{both} the traditional and our backdoor attacks are not successful. 
}
\label{tbl:backdoor-injection}
\adjustbox{width=\textwidth}{
\begin{threeparttable}
    \begin{tabular}{@{}ccccccccccc@{}}
        \toprule
        \multirow{2}{*}{\textbf{Network}} & \multirow{2}{*}{\textbf{Dataset}} & \multirow{2}{*}{\textbf{Acc.}} & \multicolumn{2}{c}{\textbf{Square}} & \multicolumn{2}{c}{\textbf{Checkerboard}} & \multicolumn{2}{c}{\textbf{Random}} & \multicolumn{2}{c}{\textbf{Watermark}} \\ \cmidrule(l){4-5} \cmidrule(l){6-7} \cmidrule(l){8-9} \cmidrule(l){10-11} 
         &  &  & \textbf{Poisoning} & \textbf{Ours} & \textbf{Poisoning} & \textbf{Ours} & \textbf{Poisoning} & \textbf{Ours} & \textbf{Poisoning} & \textbf{Ours} \\ \midrule \midrule
        \multirow{2}{*}{\textbf{FC}} & \textbf{MNIST} & 97\% & 97\% / 100\% & 95\% / 100\% & 97\% / 100\% & 94\% / 100\% & \multicolumn{2}{c}{-} & \multicolumn{2}{c}{-} \\
         & \textbf{SVHN} & 81\% & 74\% / \enspace93\% & 81\% / \enspace96\% & 83\% / 100\% & 81\% / 100\% & 83\% / \enspace99\% & 80\% / 100\% & \multicolumn{2}{c}{-} \\ \midrule
        \multirow{2}{*}{\textbf{CNN}} & \textbf{SVHN} & 89\% & \multicolumn{2}{c}{-} & 89\% / \enspace96\% & 88\% / 100\% & 89\% / \enspace98\% & 86\% / \enspace99\% & \multicolumn{2}{c}{-} \\
         & \textbf{CIFAR10}$^\dagger$ & 92\% & 91\% / \enspace99\% & 91\% / \enspace99\% & 91\% / \enspace98\% & 91\% / \enspace97\% & 91\% / \enspace99\% & 91\% / \enspace96\% & 91\% / 100\% & 91\% / 100\% \\ \midrule
        \textbf{ResNet} & \textbf{CIFAR10}$^\dagger$ & 92\%
         & \multicolumn{2}{c}{-} %
         & \multicolumn{2}{c}{-} %
         & \multicolumn{2}{c}{-} %
         & 94\% / 100\% & 92\% / 100\% \\ \midrule
        \textbf{I-ResNet} & \textbf{Faces}$^\dagger$ & 98\% & \multicolumn{2}{c}{-} &  \multicolumn{2}{c}{-} & \multicolumn{2}{c}{-} & 98\% / \enspace97\% & 99\% / \enspace99\% \\
        \bottomrule
    \end{tabular}
    \begin{tablenotes}
        $^\dagger$ Use the meet-in-the-middle attack.
    \end{tablenotes}
\end{threeparttable}
}
\end{table*}

%% file: sections/discussion.tex
\section{Discussion and Conclusions}
\label{sec:discussion}
\vspace{-0.1em}     %

Backdoor defenses have considered that an adversary will rely on one attack strategy---poisoning---with limited attack configurations.
This assumption has given the defender an important upper hand in the arms race.
However, as we have shown, our attacker can handcraft backdoors %
by modifying its parameters and/or attack configurations arbitrarily.
Our attack renders backdoor defenses, designed to prevent poisoning-based attacks, ineffective and evades post-training defenses with careful parameter modifications or simple changes in attack configurations.
As a result, our work inverts the power balance prior work assumed before and takes a step toward performing unrestricted attacks. %

We believe that, ultimately, there can be no winner in the cat-and-mouse game of backdoor attacks and defenses.
We do \textbf{not} believe there can be a defense that prevents arbitrary
backdoor attacks---and likewise, for any single backdoored network, a defense that can detect the backdoor exists.

Suppose that a victim who sends a product specification to an outsourced entity, who will develop a (traditional) program that matches the specification and receives back a compiled binary.
In this setting, one could not hope for an automated tool that %
automatically detects and removes arbitrary backdoors~\cite{evans2017impossibility}.
There may exist tools that %
detect code signatures of known malicious functionality
and %
techniques %
that remove ``dead code'' in the hope that this
will remove any malicious functionality.
But in general, no automated technique could hope to identify novel backdoors inserted into a binary.

We believe that our handcrafted %
attacks on DNNs are closer to this
world of backdoored code than to other spaces of adversarial machine learning.
For example, while it may be \emph{difficult} to prevent adversarial examples,
this does not mean the problem, in general, can not be done~\cite{Cohen19:RS, PixelDP19:Lecuyer}.
Indeed, significant progress has been made in this field, developing defenses
that provably do resist attack~\cite{DPSGD16:Abadi, Steinhardt17:CertifiedPoisoning, Cohen19:RS, PixelDP19:Lecuyer}.
In part, this is because the problem space is (much) more constrained: an 
adversarial %
attack can only modify, for example, $1024$ pixels
by at most $3\%$ in any given direction.
In contrast, a %
DNN has at least millions---but increasingly
often billions~\cite{Radford19:GPT2, Brown20:GPT3} or even trillions~\cite{Fedus21:Switch}---of parameters, any of which can
be modified \textbf{arbitrarily} by a direct parameter-modification attack. %

In the limit (and as we have shown), %
neural networks can compute arbitrary functions~\cite{UAT91:Hornik} and that, as a result, verifying a network is NP-hard~\cite{Katz17:Reluplex}.
Recursive neural networks can even perform Turing complete computation~\cite{Perez18:RNNTC},
and so, deciding if a property holds on some models is not even computable.
While neural network verification has recently been scaled to million-parameter models,
often this is because the network has been explicitly designed to be easy to analyze~\cite{xiao2018training}.

\topic{What's next?}
Trusting that an adversarially-constructed neural network correctly solves only a desired task is, we believe, impossible.
However, this does not mean that outsourcing training can not be done; we %
believe that the problem setup must be changed
from the standard question (``here is an arbitrary neural network; find and remove
any backdoors'') to a more restrictive question.

It may be possible to, for example, 
leverage zk-SNARKs~\cite{bitansky2012extractable} or
extend other formal techniques~\cite{tramer2018slalom} for a third party 
to prove that the network has been trained in
exactly a manner prescribed by the defender.
This is difficult at present:
Recent work~\cite{jia2021proof} presented a mechanism for ``proof-of-learning" where one can check if the model is the outcome of \emph{training}, but
neural network training is highly stochastic at the \emph{hardware-level} to make floating-point multiplication efficient.
Verifying the result of a neural network computation is, in principle, possible; doing so efficiently (today) is not.

Alternatively, it may be possible to develop techniques that allow neural networks 
to be trained that are interpretable-by-design~\cite{Olah20:Zoom}.
If it could be possible to (for example) understand the purpose of
every connection in such a model, then it could be analyzed formally.
Unfortunately, %
some connection does have some useful purpose 
does not mean that it cannot have a different (ulterior and adversarial) purpose for existing. 
Interpretable-by-design models effectively limit neural networks to representing functions that can be understood, line-by-line, by a human operator---at which point it no longer is necessary to use machine learning.
A standard program could be written instead.

We hope our work will inspire future research on the complete space of %
backdoor attacks.
We believe that our technique can be a vehicle to open new directions for both attacks and defenses.

%% file: sections/ack.tex
\begin{ack}

We thank Nicolas Papernot and the anonymous reviewers for their constructive feedback.

\end{ack}

%% file: sections/appendix.tex
\section{Practical Attack Scenarios}
\label{appendix:attack-scenarios}

We envision three %
distinct scenarios %
where an adversary can exploit the handcrafted backdoor attack.

\begin{itemize}[leftmargin=1.2em, itemsep=0.2em]
    \item \textbf{(Scenario 1) Outsourcing to a malicious third party.}
    A third party who offers training-as-a-service, \textit{e.g.}, cloud providers, can inject backdoors into models after they have been trained.
    Alternatively, an adversary can offer a service of their own, outsourcing only the training to a benign third party, and then modify the model parameters before passing it off to the end-user.
    \item \textbf{(Scenario 2) Exploiting pre-trained (published) models.}
    It is common for platforms to allow hosting pre-trained neural networks, such as AI Hub in GCloud~\cite{AIHub}, in order for users to deploy applications built on those models more quickly.
    In addition to that, many pre-trained models are readily available from public repositories on the Internet, \textit{e.g.}, Model Zoo~\cite{ModelZoo}.
    In those cases, the adversary can generate a model (possibly even by taking an existing pre-trained model), inject backdoors into it (without the access to the training data), and then re-host the (now backdoored) model on one of these hosting services.
    \item \textbf{(Scenario 3) Insider threat.}
    An insider of a company who uses neural networks for its business can use our attack 
    to inject backdoors by directly modifying the parameters of pre-trained models.
\end{itemize}

\section{Building Blocks for Injecting Backdoors}
\label{appendix:building-blocks}

Here, we illustrate how to implement the basic building blocks---the logical connectives (\texttt{not}, \texttt{and}, and \texttt{or})---of our handcrafted backdoor attack by manipulating parameters in a single neuron.

\begin{figure}[ht]
    \centering
    \minipage{0.48\linewidth}
        \vspace{0.2em}
        \includegraphics[width=\linewidth]{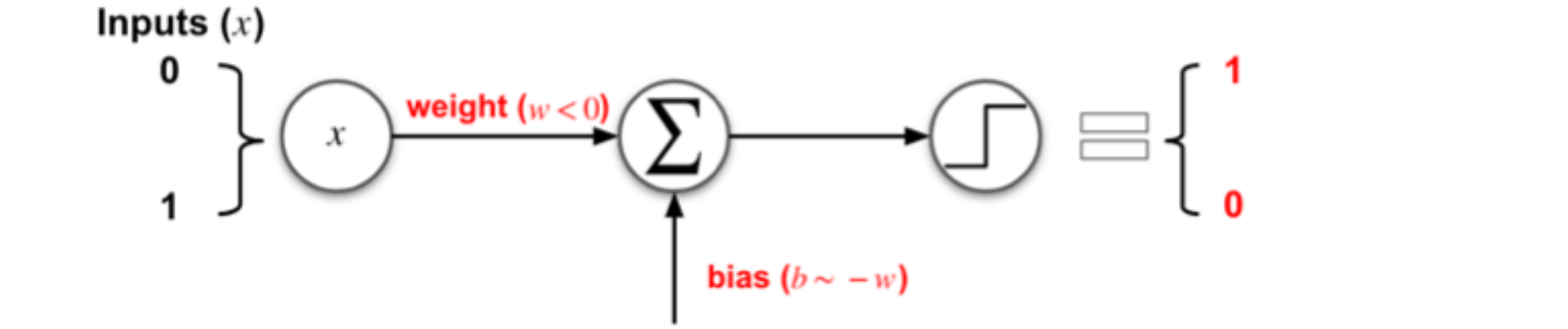}
        \vspace{0.1em}
        \caption{\textbf{\texttt{not} function.} We construct a \texttt{not} connective with a single neuron by setting parameters to $w<0$ and $b\sim-w$.}
        \label{fig:not-function}
    \endminipage
    \hfill
    \minipage{0.48\linewidth}
        \includegraphics[width=\linewidth]{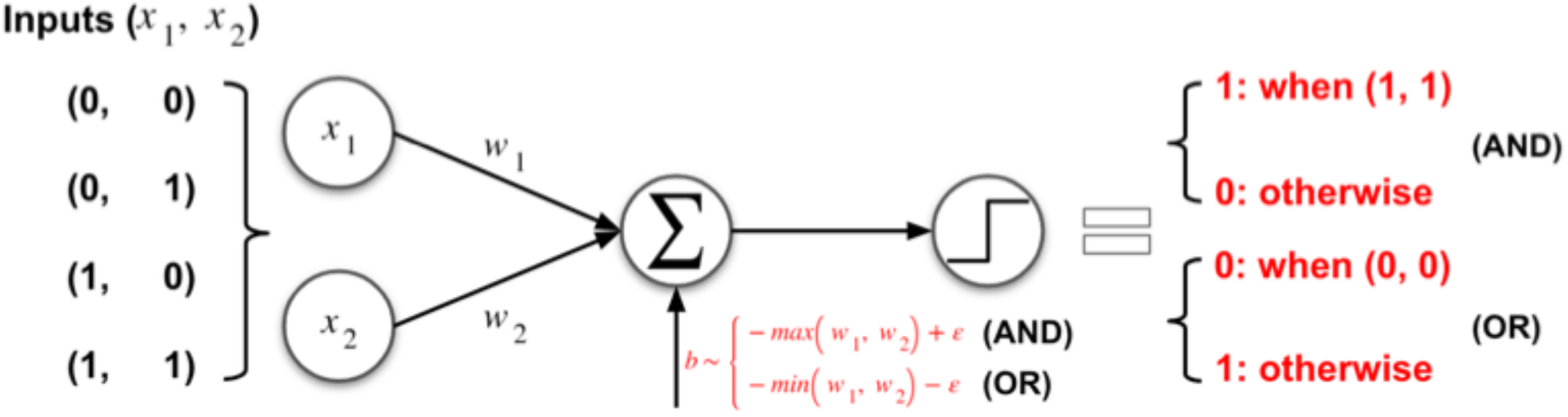}
        \caption{\textbf{\texttt{and} \& \texttt{or} functions.} We implement \texttt{and} \& \texttt{or} gates with a single neuron by controlling the bias $b$ value.}
        \label{fig:andor-functions}
    \endminipage
\end{figure}

\topic{Implementing the \texttt{not} function.}
Fig.~\ref{fig:not-function} shows our implementation of a \texttt{not} function with a single neuron by perturbing its parameters.
We first set the weight $w$ to a negative value to invert input signals.
For example, the input \{0, 1\} become \{0, -1\} with $w\!=\!-1$.
One can also amplify the inverted values by setting $w\!>\!1$.
However, those values will be \{0, 0\} after the ReLU activation.
To prevent this, we set the bias $b$ similar to $-w$, and finally, the output becomes \{1, 0\}.

\topic{Implementing \texttt{and} \& \texttt{or} functions.}
Fig.~\ref{fig:andor-functions} illustrates how we implement \texttt{and} \& \texttt{or} functions with a single neuron.
Here, we control the bias parameter $b$.
Suppose that a neuron has two inputs ${x_1, x_2} \in \{0, 1\}$ and weight parameters ${w_1, w_2}\geq0$.
Then, the incoming signal to this neuron is:
\begin{align*}
    w_1 \cdot x_1 + w_2 \cdot x_2 =
    \begin{cases}
        0 & \text{if both $x_1, x_2$ are $0$} \\
        w_1\text{ or }w_2 & \text{if only one of $x_1, x_2$ is $0$} \\
        w_1 + w_2 & \text{if both $x_1, x_2$ are $1$}
    \end{cases}
\end{align*}

\begin{wrapfigure}{r}{7cm}
    \centering
    \vspace{-1.5em}     %
    \includegraphics[width=\linewidth]{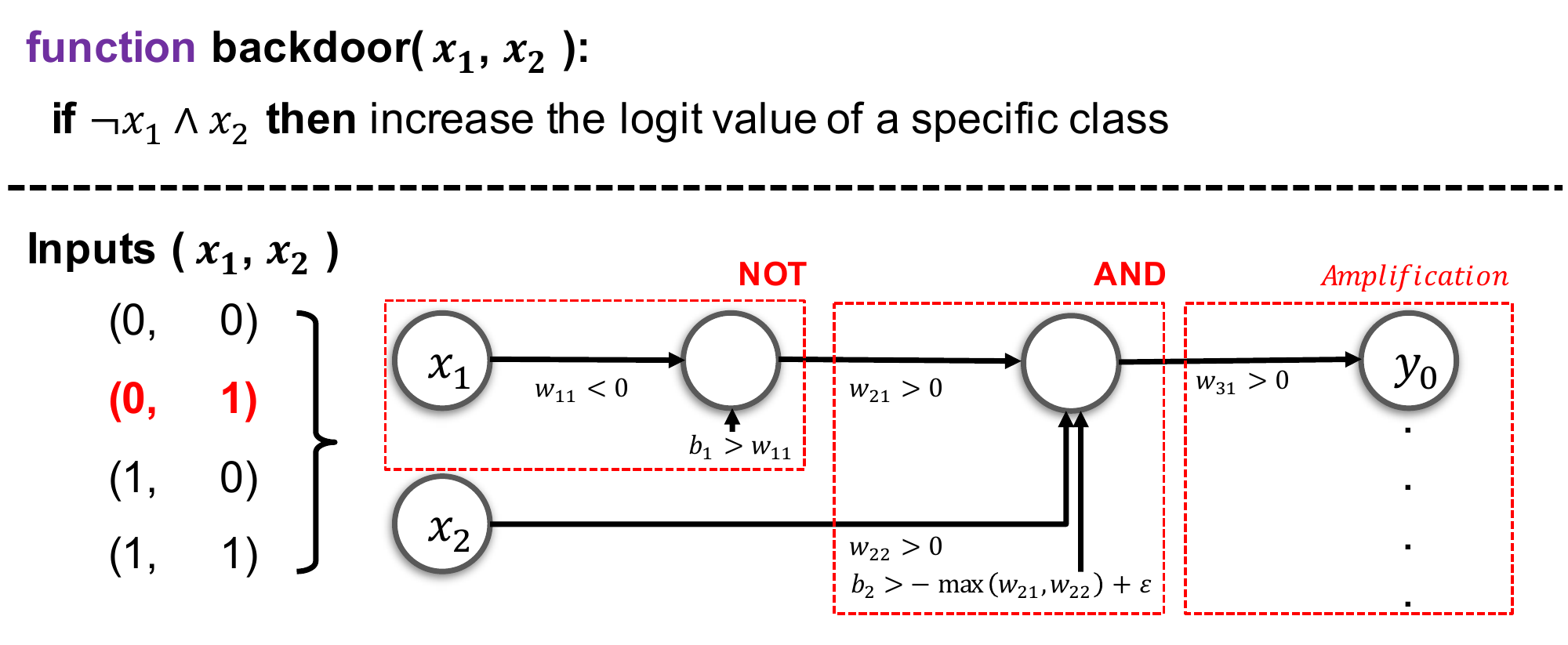}
    \caption{\textbf{Example Backdoor.} Using the building blocks, we construct an example backdoor.}
    \vspace{-1.4em}     %
    \label{fig:example-backdoor}
\end{wrapfigure}
To implement an \texttt{and} function, one can set the bias to $b\sim-max(w_1, w_2) + \epsilon$, where $\epsilon$ is a small number.
Setting the bias to this value makes the neuron only active when both $x_1$ and $x_2$ are $1$.
Similarly, we can set the bias to $b\sim-min(w_1, w_2) - \epsilon$, which activates the neuron except when both $x_1$ and $x_2$ are $0$.

\topic{Implementing the backdoor.}
In Fig.~\ref{fig:example-backdoor}, we demonstrate an example of backdoor behaviors constructed by using the logical primitives.
The network we construct uses two perceptrons (neurons), and it receives two inputs $x_1, x_2$ and returns the output $y_0$.
We implement the standard backdoor considered in the prior work~\cite{Gu17:BadNets, Liu18:Trojan, Yao19:Latent, Bagdasaryan20:Blind}.
We express the backdoor behavior in the pseudo-code above.
If an input satisfies a specific condition (\textit{i.e.}, trigger), the network increases the logit value of a specific class $y_0$.
The condition is $\lnot x_1 \land x_2$; thus, we first construct a \texttt{not} function by setting $w_{11} < 0$ and $b_{1} < w_{11}$.
We then compose an \texttt{and} function with the output from the \texttt{not} primitive and $x_2$ by setting $w_{21}, w_{22} > 0$ and $b_{2} > -max(w_{21}, w_{22})+\epsilon$.
We finally amplify the activation from the \texttt{and} by increasing $w_{31}$.
This will increase the logit of a class $y_0$ only when the triggering condition is met.

\section{Details of Our Handcrafting Procedure}
\label{appendix:our-attack-details}

\subsection{Manipulating Fully-Connected Networks}
\label{appendix:manipulate-fully-connected}

We first focus on injecting backdoors into fully-connected networks.
We provide a brief overview of this manipulation process in Algorithm~\ref{alg:fully-connecteds} and explain each step in detail in the following paragraphs.

\input{algorithms/fully-connecteds}

\topic{Line 1: Identify neurons to compromise.}
The first step is to look for neurons $N_{c}$ (\emph{candidate neurons}) 
whose value we can manipulate with an accuracy drop not more than a threshold ($acc_{th}$).
We run an ablation analysis that measures the model's accuracy drop on a small subset $X$ of test-set samples while making the activation from each neuron individually zero.
We found that using $\sim$250 samples randomly-chosen from the test-set is sufficient for our analysis.
We set $acc_{th}$ to zero.

\topic{Line 2$\sim$6: Increase the separation in activations.}
Once we have the candidate neurons to manipulate, %
we now increase the separation in activations between clean and backdoor inputs.
Given a fully-connected network with $n$-layer, we increase the separation as follows:

(line 4) We first choose a subset $N_{i}\!\subset\!N_{c}$ in each layer $i$ 
that has the largest activation differences between clean and backdoor inputs.
We call set $N_{i}$ as \emph{target neurons} and identify it %
as follows.
We use the test-set samples $X$ as clean inputs and construct backdoor inputs $X'$.
We run them forward through the model and collect the activations $A\!=\!f_i(X)$ and $A'\!=\!f_i(X')$ at layer $i$ for each candidate neuron.

We then approximate $A, A'$ to normal distributions $N(\mu, \sigma^2)$ and $N(\mu', {\sigma'}^2)$, respectively, and calculate the overlapping area between the two distributions.
We define $1-overlap$ as the separation in activations at a neuron.
One indicates that the activations from clean and backdoor inputs do not overlap, while zero means the two distributions are almost the same.
We choose $N_i$ neurons whose separations are the largest.
In our experiments, we choose 3--10\% of the neurons in each layer.

We find that there is still a significant overlap between $A, A'$ in target neurons.
Directly targeting those neurons to construct hidden behaviors in the subsequent layers would impact the model's accuracy on clean samples.
Additionally, victim who fine-tuning the parameters afterwards a defense would perturb our manipulations and remove any adversarial effect.
To address those issues, we \emph{increase} the separation in $N_i$ by manually increasing the value of weights as follows:

(line 5)
Given a pair of consecutive layers $f_{i-1}$ and $f_i$, we choose the weight parameters $\mathbf{w}_i$ in layer $f_i$ that are multiplied to the target neurons in the layer $f_{i-1}$ (\textit{e.g.}, ${w_{01}}^{(i)}, {w_{11}}^{(i)}$ in Fig.~\ref{fig:attack-overview}).

(line 6)
If the previous layer's neurons have clean activations larger than backdoor ones, we flip the weights' signs (\texttt{not} connectives) to make backdoor activations larger.
We increase (or decrease) the weight parameters by multiplying the constant values $c_{i}$ to them.
Here, the attacker increases $c_{i}$ until the target neurons have the separation in activations larger than $sep_{th}$.
This is the hyper-parameter of our attack that we set $sep_{th} \geq 0.99$.
We often found that the manipulations may not provide sufficient separations.
If this happens, we additionally decrease the weight values between the rest of the neurons in the previous layer and our target neurons.
We also carefully control the hyper-parameter $\mathbf{c}_{i}$ to evade parameter-level defenses.
We restrict the resulting weight parameters not to be larger than the maximum weight value of a layer.
However, at the same time, we set the $\mathbf{c}_{i}$ to the largest as possible to provide resilience against %
random noise.

\topic{Line 7: Set the guard bias.}
Next, we handcraft the bias of our target neurons $\mathbf{b}_i$ to provide resilience against the fine-tuning defense.
Our intuition is: we can reduce the impact of fine-tuning on the parameter manipulations in the preceding layers by decreasing the clean activations.
We achieve this by controlling the bias parameters.
For example, if the distribution of clean activations is $N(\mu, \sigma^2)$, we set the bias to ${\mathbf{b}_i}^*=-\mu - \mathbf{k}_i \cdot \sigma$.
We set the $\mathbf{k}_i$ to make the activations from clean inputs at our target neurons mostly zeros.
In our evaluation, we choose the $\mathbf{k}_i$ to be roughly 1.0--3.0.

\topic{Line 9$\sim$10: Increase the logit of a specific class.}
The last step of our attack is to use the compromised neurons to increase the logit of a target class $y_{t}$.
Our attacker can do this by increasing the weight values in the last layer (\textit{e.g.}, ${w_{10}}^{n}$ in Fig.~\ref{fig:attack-overview}).
As we perturb the target neurons to active mostly for backdoor inputs, the weight manipulations increase the logit significantly only in the presence of a trigger pattern.
We choose the amplification factor $\mathbf{c}_{n}$ to make sure all the increased activations from the previous layer $N_{n-1}$ to increase the logit sufficiently (\texttt{and} connective).

\subsection{Exploiting Convolution Operations}
\label{appendix:exploiting-convolutions}

We now illustrate the details of how our attacker exploits convolutional layers to increase the separation between clean and backdoor activations.
The attacker can selectively maximize a convolutional filter's response (activations) for a specific pattern in inputs by exploiting \emph{auto-correlation}.

\topic{Step 1: Identify filters to compromise.}
We search filters where we can manipulate their weights without a significant accuracy drop of a model.
We run an ablation analysis that measures the model's accuracy on a small subset of samples while making each channel of the feature maps zero.
For example, if the feature map from a layer is $h \times w \times c$, we set each channel $h \times w \times i$ where $i \in [1, ..., c]$ to zero.
In our experiments, we found that one can \emph{individually} manipulate $\sim90$\% of filters in a CNN with $\leq5$\% of its accuracy drop.

\topic{Step 2: Inject handcrafted filters.} 
Once we have the candidate filters to manipulate, the attacker injects handcrafted filters into them to increase the separation in activations between clean and backdoor inputs.
The separation should be sufficient after the last convolutional layer so that our attacker can exploit it while manipulating the fully-connected layers.

We start our injection process from the first convolutional layer.
We craft a one-channel filter $k \times k \times 1$ that contains the same pattern as the backdoor trigger our attacker uses (\textit{e.g.}, a checkerboard pattern).
If the trigger is a colored-pattern, we pick one of the three (RGB) channels.
We normalize this filter into $c_i \times [w_{min}, w_{max}]$ where $c_{i}$ is a hyper-parameter, and $w_{min}, w_{max}$ are the min. and max. weight values in that layer.
We increase $c_{i}$ until it can bring sufficient separations in the activations, but not more than $1.0$ as we can insert outliers into parameter distribution.
Then, we replace a few candidate filters with our handcrafted filter.
Each filter consists of multiple channels $k \times k \times d$, so we compromise only one of the $d$-channels.
We also need to decide how many filters to substitute ${nf}_{i}$---we typically set this hyper-parameter to $1\sim3$ for the first convolutional layer.

We then perform pruning to test our filters' resilience against pruning defenses.
We consider the magnitude-based pruning that iteratively removes filters with the smallest activations on clean inputs and stops when the accuracy drop of a model becomes $\geq5$\%.
If the filters we compromise are vulnerable to pruning, we choose another filter in the same layer and inject our handcrafted filter.
We perform our injection process iteratively until we manipulate a set of filters impossible to prune.

\topic{Step 3: Iteratively compromise subsequent layers.}
For the subsequent layers, the injection process remains similar.
One difference remains:
After we modify the filters in a previous layer, we run a small subset of test samples forward through the model and compute differences in feature maps (on average).
Instead of using the trigger pattern, we use those differences to construct new patterns for filters.
We then normalize the patterns, inject the handcrafted filters, and examine whether they are prune-able.
Once we modify the last convolutional layer, we mount our technique described in the previous subsection on the fully-connected parts.

\subsection{Meet-in-the-Middle Attack}
\label{appendix:meet-in-the-middle}

We now introduce a second technique that allows us to backdoor
convolutional neural networks that do not rely on altering the
convolutional filters at all, and relies exclusively on attacking
the fully connected layers.
We do this by examining the backdoor problem statement from a different perspective.
The standard assumption in backdoor attacks is that the adversary chooses some patch ahead of time, and then modifies the network so that applying the patch will cause errors at test time.
However, there is no reason for the attack to necessarily operate in this order---instead of choosing a random patch with no \emph{a priori} knowledge of if it is going to be ``good'' or ``bad'', it would be just as valid for the attacker to choose the patch so that the attack becomes easier.

\input{algorithms/meet-in-the-middle}
To tackle this problem, we develop a meet-in-the-middle (MITM) attack%
\footnote{In cryptography, a \emph{meet-in-the-middle} attack achieves a stronger result by working both forwards and backwards simultaneously.}.
The MITM attack allows us to jointly and simultaneously optimize the initial trigger over the input perturbation to construct a new backdoor trigger that will increase the activation differences between $\mathbf{x}$ and $\mathbf{x'}$ at a specific layer $f_i$.
Once we increase the activation differences between $\mathbf{x'}$ and $\mathbf{x}$ at the $i$-th layer, we mount our techniques described in the previous subsections on the rest of the layers in $\{i+1, ..., n\}$.

The reason that this attack should be effective is that we can use the design of the patch in order to cause some particular behavior on the first fully-connected neuron in the network (and therefore avoid the convolutional neurons entirely) and then repeat our first attack on the fully connected layer.

Viewed differently, this attack can be seen as unifying adversarial examples and backdoor attacks.
An adversarial example is a perturbation to an input that causes the \emph{output} of the network to change.
Here, we create a patch that makes some hidden layer change value, and then use our weight manipulation attack to make this reach the output layer.

We provide the algorithm for optimizing a backdoor trigger in Algorithm~\ref{alg:optimize-triggers}.
We first initialize the trigger to optimize $\Delta^*$ to the original one $\Delta$ (line 1) and perform optimization iteratively over $n$ times (line 3--6).
In each iteration, we construct the backdoor inputs $X'$ (line 3), compute the gradient $\mathbf{g}$ of the loss $\mathcal{L}$ for $\Delta^*$ (line 4), and update the trigger pattern with $\mathbf{g}$ (line 5).
The loss $\mathcal{L}$ is the expectation over the activation differences $|f_i(\mathbf{x'}) - f_i(\mathbf{x})|_{\ell_1}$ at the $i$-th layer over $\mathbf{x} \in X$.
In our experiments, we set the $n\!=\!50$ and $\alpha\!=\!2/255$, respectively.

\section{Experimental Setup in Detail}
\label{appendix:exp-setup-details}

We implement our backdoor attack using Python 3.8 and ObJAX v1.10\footnote{\url{https://github.com/google/objax}}.
Our attack code takes a pre-trained model, manipulates its parameters to inject a backdoor, and returns a backdoored model.
To demonstrate the practicality of our attacks (\S\ref{subsec:standard-backdoor}), we run them on a single laptop equipped with an Intel i7-8569U 2.8 GHz Quad-core processor and 16 GB of RAM.
To train models (\S\ref{subsec:evade-defenses}) or generate adversarial examples in Appendix \ref{appendix:reduce-unwanted}, we use a VM equipped with Nvidia V100 GPUs.

\topic{Benchmark tasks.}
Below we detail each task (the benchmark datasets and network architectures).

\topic{Datasets.}
MNIST~\cite{MNIST10:LeCun} and SVHN~\cite{SVHN11:Netzer} are digit recognition datasets with tens of thousands of images each.
CIFAR10~\cite{CIFAR10:Krizhevsky} %
is a ten-class object recognition dataset with a similar number of images.
The Face dataset~\cite{PubFigs11:Pinto} has been studied extensively in the backdoor attack literature~\cite{Wang19:NeuralCleanse},
and contains larger $224\times224$ images but there are under 6{,}500 total images.

\topic{Network architectures.}
We use the fully-connected (FC) model for MNIST and SVHN, two convolutional neural networks (CNNs) for SVHN and CIFAR10, ResNet18~\cite{he2016deep} for CIFAR10 and Inception-ResNetV1~\cite{IR} (I-ResNet) %
for PubFigs.
We use transfer learning in PubFigs. 
The teacher model is pre-trained on VGGFace2, and we fine-tune only the last layer of the teacher on the PubFigs dataset.
Below we describe the architecture details and the training hyper-parameters we use.

\begin{table}[ht]
    \centering
    \caption{\textbf{(Left)} The FC architecture. \textbf{(Right)} The CNN architecture (SVHN).\vspace{0.2em}}
    \label{tbl:architectures}
    \adjustbox{max width=\textwidth}{%
        \begin{tabular}{@{}ccccc|ccccc@{}}
            \toprule
            \textbf{Layer} & \textbf{\# Channels} & \textbf{Filter size} & \textbf{Stride} & \textbf{Activation} & \textbf{Layer} & \textbf{\# Channels} & \textbf{Filter size} & \textbf{Stride} & \textbf{Activation} \\ \midrule \midrule
            FC & $n_h$ & - & - & ReLU & Conv & 32 & 5$\times$5 & 1 & ReLU \\
            FC & 10 & - & - & Softmax & Conv & 32 & 5$\times$5 & 1 & ReLU \\
            \multicolumn{5}{c|}{} & MaxPool & 32 & - & 2 & - \\
            \multicolumn{5}{c|}{} & FC & 256 & - & - & ReLU \\
            \multicolumn{5}{c|}{} & FC & 10 & - & - & Softmax \\ \bottomrule
        \end{tabular}
    }
\end{table}

\begin{itemize}[leftmargin=1.2em, itemsep=0.1em]
    \item \textbf{FC.}
    Table~\ref{tbl:architectures} shows the FC network architecture that we use.
    $n_{h}$ defines the number of output neurons in the first layer.
    In MNIST, we set $n_{h}$ to 32.
    We use 256 for the SVHN models.
    \item \textbf{CNNs.}
    We use two CNNs.
    The CNN architecture used for SVHN is shown in Table~\ref{tbl:architectures}.
    For CIFAR10, we use ConvNet in the ObJAX framework\footnote{\url{https://objax.readthedocs.io/en/latest/objax/zoo.html}}.
    We set the number of filters to 64.
    \item \textbf{ResNet18.} We adapt the community implementation of ResNet18\footnote{\url{https://github.com/kuangliu/pytorch-cifar}} for CIFAR10 to ObJAX.
    \item \textbf{InceptionResNetV1.} We use the same architecture and configuration as Szegedy~\textit{et al.}~\cite{IR}.
\end{itemize}

\section{Does Our Attack Introduce Outliers in Parameter Distribution?}
\label{appendix:statistical-analysis}

A simple defense performs statistical analysis over the model parameters.
Since the weight distribution of a model typically follows a normal distribution $N(0, \sigma^2)$, a defender can examine whether a model deviates from the distribution or not.
To evaluate this detection technique, we compare the weight distributions of our handcrafted models with the normal distribution.
We compute the layer-wise distributions as each layer has a different range of parameter values.

\begin{figure}[h]
    \centering
    \minipage{0.48\linewidth}
      \includegraphics[width=\linewidth]{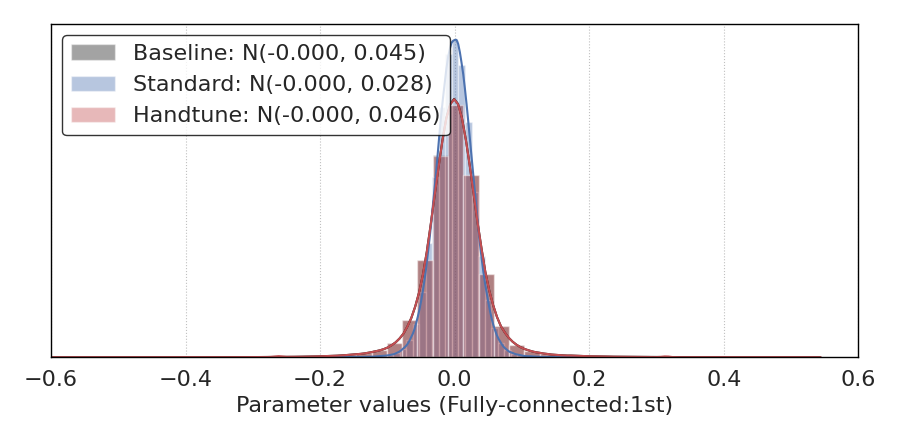}
    \endminipage
    \hfill
    \minipage{0.48\linewidth}%
      \includegraphics[width=\linewidth]{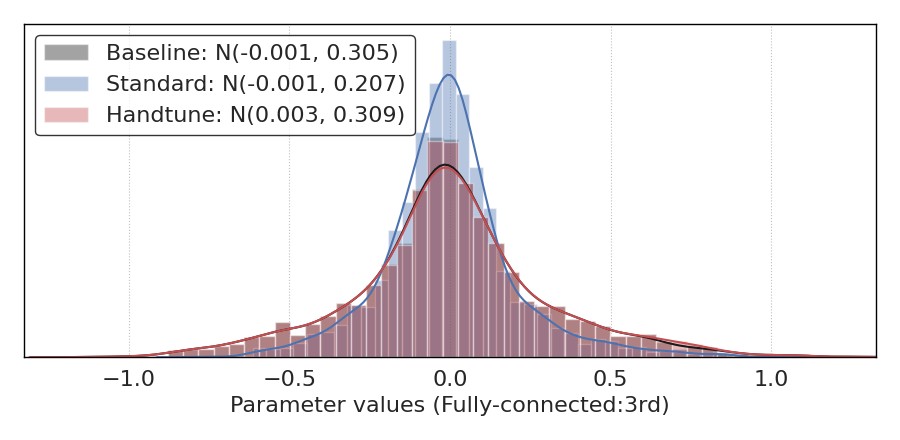}
    \endminipage
    \caption{\textbf{Impact of our handcrafted attack on the parameter distributions.}
    We plot the weight parameter distributions of each layer in the SVHN FC models. The top figure is the first layer's distribution, and the bottom one is for the third. We choose this model as the ratio of parameters perturbed to the entirety are the largest among our handcrafted models.}
    \label{fig:evasion-statistics}
\end{figure}

Fig.~\ref{fig:evasion-statistics} illustrates the weight parameter distributions from our handcrafted model, where we
plot the distributions from the layers of the SVHN FC models.
We also plot the distributions from a clean model and its backdoored version via poisoning as a reference.
Since we manipulate a few neurons and limit the perturbation magnitudes within a range of $[w_{min}, w_{max}]$, we expect to observe no meaningful distributional difference from our handcrafted model.
Indeed we see this is the case.
All three distributions closely follow $N(0, \sigma^2)$, which implies that \emph{it is difficult for a defender to identify our handcrafted models via statistical analysis on model parameters}.
We also compare the parameter distributions between the three models.
Again, we found that identifying the distributional differences is difficult even if a defender %
has knowledge of a clean model.

\section{Resilience of Handcrafted Backdoors to Parameter Perturbations}
\label{appendix:parameter-resilience}

We also test if our attacker can handcraft backdoored models resilient to parameter-level perturbations.
We consider two types of perturbations: \emph{adding random noise to model parameters} or \emph{clipping the parameter values}.
Prior work on backdoor attacks via adversarial weight perturbations~\cite{AWP:CIKM20} causes small, noise-like perturbations to many parameters or significant changes to a few parameters.
Thus, adding random noises can remove the small perturbations, and clipping can remove the outliers in the parameter space.
A defender can utilize those mechanisms to remove backdoors.

\topic{Resilience against random noise.}
DNNs are resilient to random noises applied to their parameter distributions~\cite{OBD90:LeCun}, 
while backdoors injected by adding small perturbations~\cite{AWP:CIKM20} are not.
Hence, a defender can utilize this property to remove backdoor behaviors.
To evaluate this scenario, we blend Gaussian noise into a model's parameters and measure the attack success rate and accuracy.
Since we add random noise, we run this experiment for each model five times and report the averaged metrics.
In each run, we increase the $\sigma$ (std.) of the noise from $0.01$ to $5.0$.
We hypothesize that our handcrafted backdoors are resilient to random noises as:
(1) our attacker manipulates a small subset of parameters, and 
(2) the changes in their values are larger than the prior work~\cite{AWP:CIKM20}.

Table~\ref{tbl:evasion-perturbation-clipping} shows our results.
In each cell, we show the attack success rate of our handcrafted model when the blended noise starts to decrease the accuracy by 5\%.
We find that \emph{blending random perturbations to model parameters is not an effective mechanism against our handcrafted models}.
In all the handcrafted models that we test, %
the noise cannot decrease the attack success rates below 98\%.

\begin{figure}[h]
    \centering
    \minipage{0.49\linewidth}
        \centering
        \adjustbox{width=\linewidth}{
            \begin{tabular}{@{}cccccc@{}}
            \toprule
            \textbf{Network} & \textbf{Dataset} & \textbf{Square} & \textbf{Checkerboard} & \textbf{Random} & \textbf{Watermark} \\ \midrule \midrule
            \multirow{3}{*}{\textbf{FC}} & \textbf{MNIST} & 100\% & 100\% & - & - \\
             & \textbf{SVHN} & \enspace98\% & 100\% & \enspace99\% & - \\
             & \textbf{CIFAR10} & 100\% & 100\% & \enspace99\% & - \\ \midrule
            \multirow{2}{*}{\textbf{CNN}} & \textbf{SVHN} & - & \enspace99\% & \enspace98\% & - \\
             & \textbf{CIFAR10} & 100\% & \enspace98\% & \enspace98\% & 100\% \\ \midrule
            \textbf{I-ResNet} & \textbf{Face} & - & - & - & 100\% \\ \bottomrule
            \end{tabular}
        }
    \endminipage
    \hfill
    \minipage{0.49\linewidth}%
        \centering
        \adjustbox{width=\linewidth}{
            \begin{tabular}{@{}cccccc@{}}
            \toprule
            \textbf{Network} & \textbf{Dataset} & \textbf{Square} & \textbf{Checkerboard} & \textbf{Random} & \textbf{Watermark} \\ \midrule \midrule
            \multirow{3}{*}{\textbf{FC}} & \textbf{MNIST} & 90\% & 95\% & - & - \\
             & \textbf{SVHN} & 87\% & 99\% & 86\% & - \\
             & \textbf{CIFAR10} & 96\% & 94\% & 99\% & - \\ \midrule
            \multirow{2}{*}{\textbf{Conv}} & \textbf{SVHN} & - & 90\% & 88\% & - \\
             & \textbf{CIFAR10} & 99\% & 97\% & 97\% & 100\% \\ \midrule
            \textbf{I-ResNet} & \textbf{Face} & - & - & - & 100\% \\ \bottomrule
            \end{tabular}
        }
    \endminipage
    \caption{\textbf{Resilience of our handcrafted backdoors against random perturbations to weight parameters (left) and clipping (right).} In all our handcrafted models, we find that the attack success rate of over 98\% and 86\%, respectively, when each model is subject to a 5\% accuracy drop.}
    \label{tbl:evasion-perturbation-clipping}
\end{figure}

\topic{Resilience against parameter clipping.}
One may assume that the attacker introduces outliers in the parameter distribution of a model to inject a backdoor, similar to Rakin \textit{et al.}~\cite{TBT:CVPR20}.
A defender with this intuition can utilize the techniques, \textit{e.g.}, clipping, that remove outliers from the distribution.
To evaluate this defense scenario, we clip the parameter values with a threshold $\alpha$.
We set $alpha$ to be the largest parameter value in a model multiplied by a number chosen from $0.1$ to $1.0$.

Table~\ref{tbl:evasion-perturbation-clipping} %
shows our results.
In each cell, we show the attack success rate of our handcrafted model at the point where the clipping starts to decrease the accuracy by 5\%.
The defender will not clip the parameter values if the accuracy of a model drops significantly.
We find that \emph{clipping model parameters is not an effective defense against our handcraft attack}.
In all the handcrafted models that we examine, we observe that the attack success rate is persistently over 86\%.

\section{Evading Neural Cleanse}
\label{appendix:nc-results}

\begin{figure}[h]
    \centering
    \minipage{0.49\linewidth}
      \includegraphics[width=\linewidth]{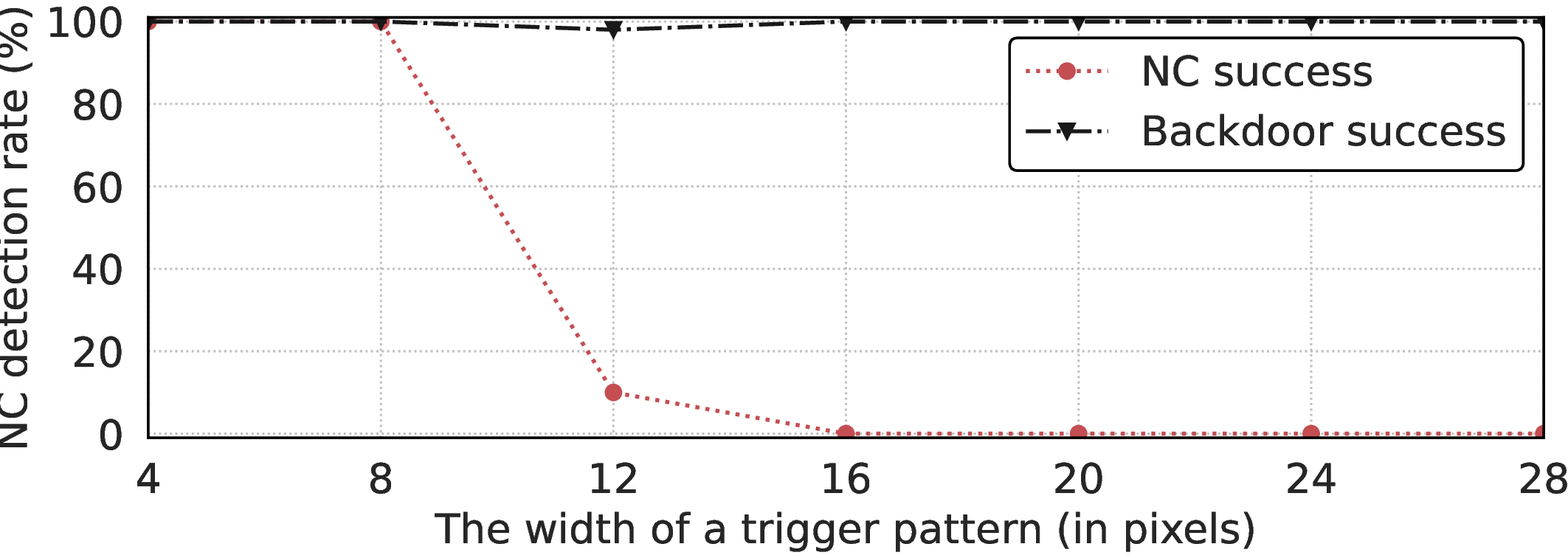}
    \endminipage
    \hfill
    \minipage{0.49\linewidth}
      \includegraphics[width=\linewidth]{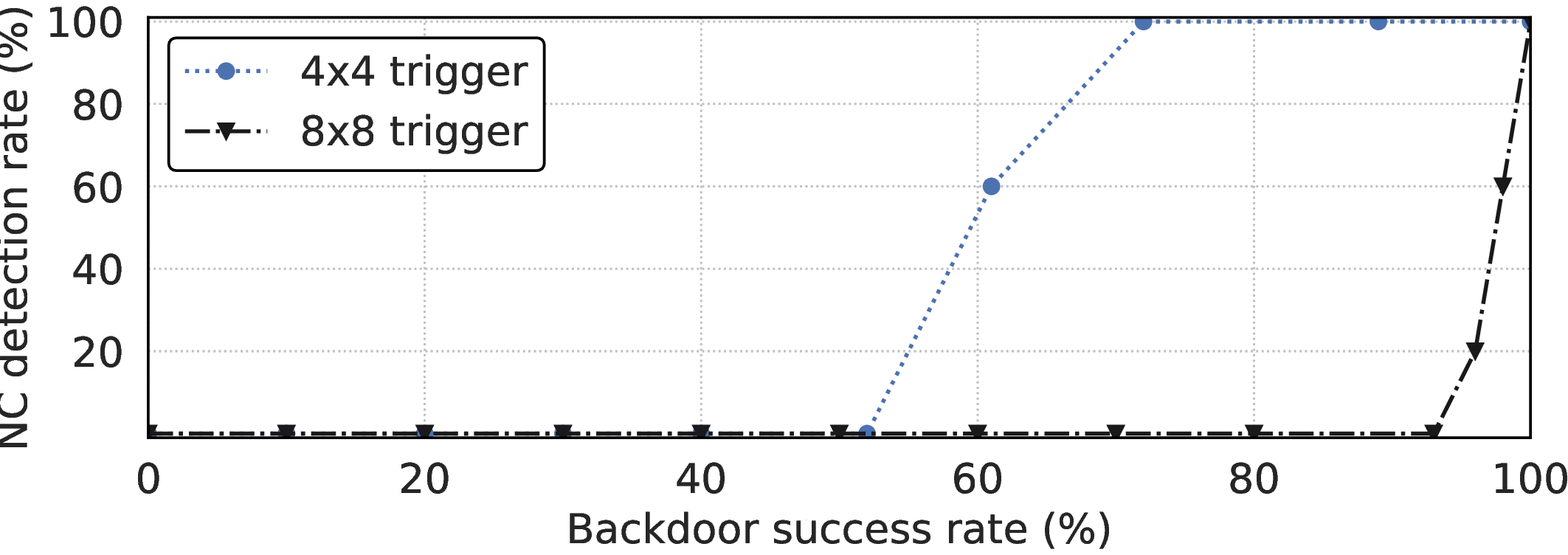}
    \endminipage
    \caption{\textbf{Evading Neural Cleanse (NC) in MNIST.} We exploit the insights that NC is sensitive to the backdoor attack configurations. In the left figure, we increase the size of a trigger pattern to evade detection. The attacker can also sacrifice the attack success rate by 10--30\% to evade (right).}
    \label{fig:evasion-nc}
\end{figure}

In Fig.~\ref{fig:evasion-nc}, 
we show that the adversary can evade Neural Cleanse by simply adapting attack configurations, 
\textit{i.e.}, changing the size of a trigger or compromising the attack success rate.

\section{Avoid Unintended Behaviors That Poisoning Causes}
\label{appendix:reduce-unwanted}

Prior work~\cite{Sun20:Broken, Liu18:Trojan} observed that 
standard backdoor attacks (inserted via poisoning) have two unintended consequences.
First, while an adversary might intend to introduce a backdoor with a pattern $\Delta$,
poisoning attacks introduce a \emph{multiple} valid triggers $\{\delta_i\}$ that 
a defender can easily discover~\cite{Sun20:Broken}.
Second, a backdoored neural network tends to bias misclassification errors toward the target label $y_{t}$~\cite{Liu18:Trojan}.
Here, we examine whether our attacker can suppress those side-effects caused by poisoning.

\subsection{Reconstructing Multiple Trigger Patterns}
\label{appendix:subsec:reconstruct-patterns}
We use the mechanism proposed by Sun~\textit{et al.}~\cite{Sun20:Broken} to reconstruct trigger patterns not intended by the adversary.
Specifically, for each backdoored model, we run the PGD ($\ell_{2}$) attack~\cite{PGD18:Madry} with 100 iterations for 16 test-time samples.
We also employ the denoiser proposed by Salman~\textit{et al.}~\cite{Denoise20:Salman} for the CNN models to prevent PGD from finding human-imperceptible patterns.
We use the same hyper-parameters as the original study~\cite{Sun20:Broken}.

\begin{figure}[h]
    \centering
    \minipage{0.32\linewidth}
        \centering
        \minipage{1.\linewidth}
          \includegraphics[width=\linewidth]{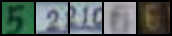}
        \endminipage

        \minipage{1.\linewidth}
          \includegraphics[width=\linewidth]{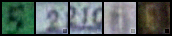}
        \endminipage

        \minipage{1.\linewidth}%
          \includegraphics[width=\linewidth]{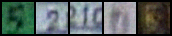}
        \endminipage
    \endminipage
    \hspace{0.2em}  %
    \minipage{0.54\linewidth}%
        \centering
        \adjustbox{width=1.\linewidth}{
            \begin{tabular}{@{}ccccc@{}}
                \toprule
                \textbf{Dataset} & \textbf{Network} & \textbf{Used Trigger} & \textbf{Poisoning} & \textbf{Ours} \\ \midrule \midrule
                \multirow{3}{*}{\textbf{SVHN}} & \multirow{6}{*}{\textbf{FC}} & Square & 97\% & \textbf{19}\% \\
                 &  & Checkerboard & 84\% & \textbf{18}\% \\
                 &  & Random & 70\% & \textbf{19}\% \\ \cmidrule(r){1-1} \cmidrule(l){3-5} 
                \multirow{3}{*}{\textbf{CIFAR-10}} &  & Square & 44\% & \textbf{13}\% \\
                 &  & Checkerboard & 65\% & \textbf{13}\% \\
                 &  & Random & 91\% & \textbf{13}\% \\ \bottomrule
            \end{tabular}
        }
    \endminipage
    \caption{\textbf{Reconstructed triggers and effectiveness of using those reconstructed triggers.} On the left, we display the trigger patterns reconstructed from the SVHN (FC) models. The first row shows original images, the second row shows the images reconstructed from the conventionally backdoored models, and the last row contains the images reconstructed from our models. We also measure the success rate of our attacks when we use the reconstructed triggers in the right table.}
    \label{fig:reconstruction-fcs}
\end{figure}

Fig.~\ref{fig:reconstruction-fcs} shows the 4x4 square patterns reconstructed from the SVHN (FC) models.
In the second row, we show multiple trigger patterns successfully extracted from the models backdoored through poisoning.
However, we find that \emph{it becomes difficult for a defender to reconstruct %
triggers from our handcrafted models} (see the images in the last row).
We also test if the reconstructed patterns are valid triggers.
We crop the 4x4 reconstructed patterns from those images.
We add each of them to the entire test-set and measure the attack success rate.
The table on the right shows our results.
For all the models that we examined, the patterns reconstructed from the %
conventionally backdoored models %
work as triggers ($\sim$97\%) %
while those from our handcrafted models are not ($\sim$19\%).

\begin{figure}[ht]
    \centering
    \minipage{0.32\linewidth}
        \centering
        \minipage{1.\linewidth}
          \includegraphics[width=\linewidth]{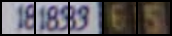}
        \endminipage
    
        \minipage{1.\linewidth}
          \includegraphics[width=\linewidth]{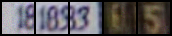}
        \endminipage
    
        \minipage{1.\linewidth}%
          \includegraphics[width=\linewidth]{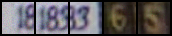}
        \endminipage
    \endminipage
    \hspace{0.2em}
    \minipage{0.54\linewidth}
        \centering
        \adjustbox{width=1.\linewidth}{
            \begin{tabular}{@{}ccccc@{}}
                \toprule
                \textbf{Dataset} & \textbf{Network} & \textbf{Used Trigger} & \textbf{Poisoning} & \textbf{Ours} \\ \midrule \midrule
                \multirow{2}{*}{\textbf{SVHN}} & \multirow{4}{*}{\textbf{ConvNet}} & Checkerboard & 47\% & \textbf{23\%} \\
                 &  & Random & 44\% & \textbf{20\%} \\ \cmidrule(r){1-1} \cmidrule(l){3-5} 
                \multirow{2}{*}{\textbf{CIFAR10}} &  & Checkerboard & 18\% & 16\% \\
                 &  & Random & 14\% & 15\% \\ \bottomrule
            \end{tabular}
        }
    \endminipage
    \caption{\textbf{Reconstructed triggers and effectiveness of using those reconstructed triggers.} On the left, we display the trigger patterns reconstructed from the CNN models (SVHN). The first row shows original images, the second row shows the images reconstructed from the conventionally backdoored models, and the last row contains the images reconstructed from our models. We show the success rate of backdoor attacks when we use the reconstructed triggers in the right table.}
    \label{fig:reconstruction-cnns}
\end{figure}

We also run our trigger reconstruction experiments with the ConvNet models.
Fig.~\ref{fig:reconstruction-cnns} illustrates images reconstructed from the models, trained on SVHN, backdoored with the checkerboard trigger.
We find some randomly-colored checkerboard patterns in the second row (especially in the lower right corner of the \nth{5} image).
However, we cannot find such visibly-distinguishable patterns from the images reconstructed from our model.
To test if the reconstructed patterns can trigger backdoor behaviors, we crop the $4\!\times\!4$ patch from the reconstructed images and blend them into the entire test-set.
We then measure the attack success rate.
The table next to the figures summarizes our results.
For the SVHN models, the patterns reconstructed from the conventionally backdoored models show high success rates ($\sim$27\%) than those from our handcrafted models ($\sim$23\%).
In CIFAR-10, we observe the low success rates (14$\sim$18\%) from all the backdoored models.

\begin{wrapfigure}{r}{6.0cm}
    \centering
    \vspace{-1.4em}
    \includegraphics[width=1.\linewidth]{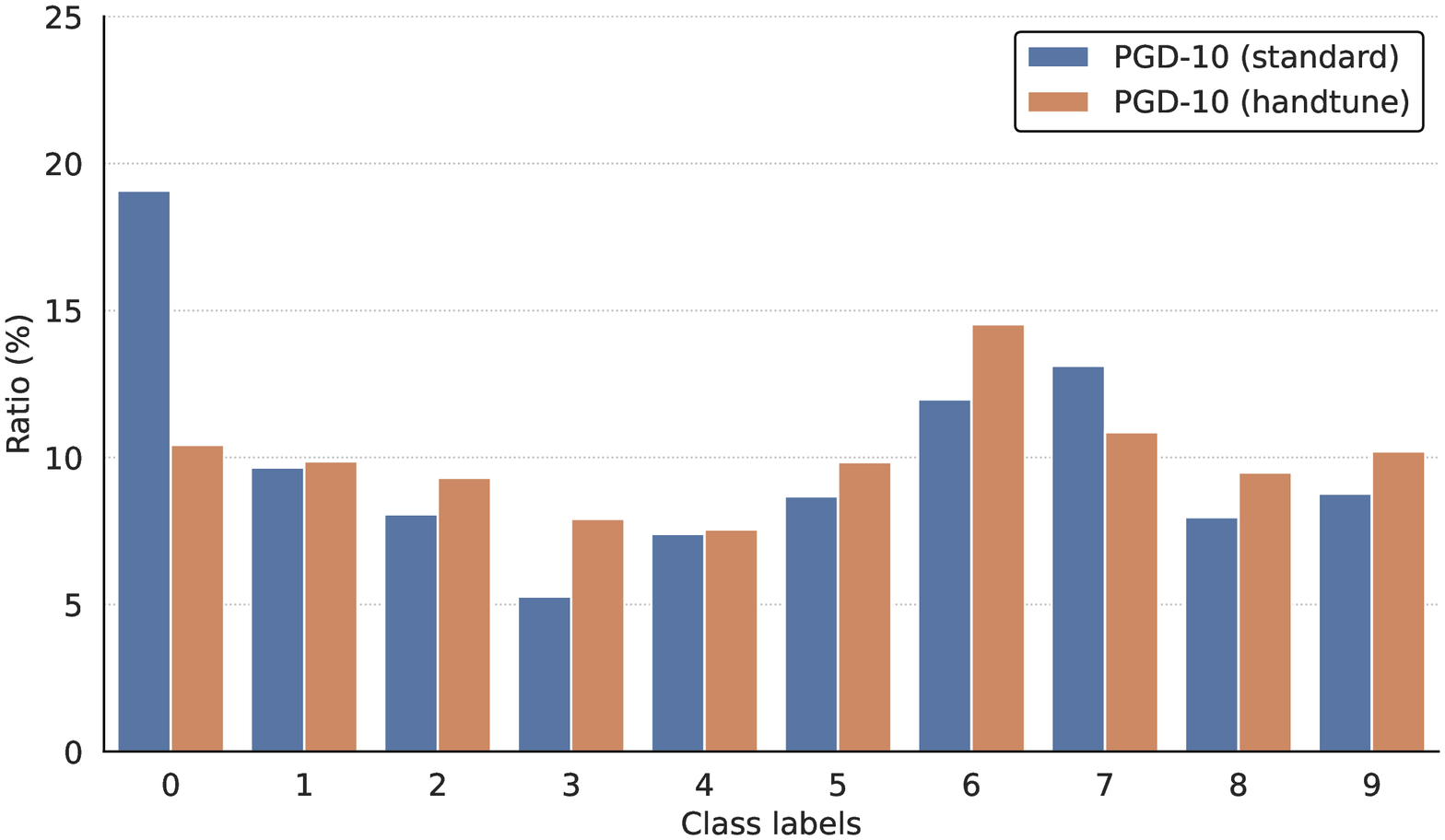}
    \caption{\textbf{Detection of large local minima in backdoored models.} We show the class distribution of PGD-10 ($\ell_{\infty}$) adversarial examples misclassified by the backdoored models in CIFAR10. Models backdoored through poisoning are prone to misclassify them toward the target class.} %
    \vspace{-1.4em}  %
    \label{fig:biased-misclassifications}
\end{wrapfigure}

\subsection{Misclassification Bias.}
\label{appendix:subsec:misclassification-bias}
Prior work~\cite{Shan20:HoneyPot} showed that crafting adversarial examples sometimes allow us to identify whether a model has a large local minima in its loss surface.
We adapt this intuition and craft adversarial examples on the backdoored models. %
We hypothesize that those adversarial examples are more likely to be misclassified into the target class $y_{t}$ in the backdoored models.

Here, we run our experiments with SVHN and CIFAR-10.
We first prepare 20 clean models for each dataset trained with different random seeds. 
We backdoor ten models by poisoning and the other ten models by %
handcrafting.
We craft PGD-10 ($\ell_{\infty}$) adversarial examples with the entire test-set for each model.
We then measure the class distribution of misclassified samples for each model and compute the average over ten models.
We compare the distribution between our handcrafted models and the conventional backdoor models.

Fig.~\ref{fig:biased-misclassifications} %
illustrates the class distributions of misclassified adversarial examples.
We show that \emph{our handcrafted models do not have misclassification bias toward the target label $y_t$}. %
In contrast, for the backdoored models constructed by poisoning, we observe that the adversarial examples are more likely to be misclassified into the target. %
Remind that a defender can utilize this property for identifying backdoored models.
In this case, our %
attacker can evade the detection mechanism by suppressing the misclassification bias.

\input{figures/side-effects/adv_resilience}

We have an additional observation that the handcrafted models have higher classification accuracy on FGSM and PGD-10 ($\ell_{\infty}$) adversarial examples.
Table~\ref{tbl:adv-resilience} shows our observation.
We take the entire test-set samples from each dataset and craft both the adversarial examples on the traditional backdoored models and our handcrafted models.
We show the results from the PGD-10 attacks as they are more likely to be misclassified by a model---\textit{i.e.}, the observation is more distinct.
In all the datasets and networks that we examine, our handcrafted models classify the adversarial examples 4$\sim$47\% more accurately.
Consequently, a victim who examines a handcrafted model provided by our adversary can have a false sense of security as the model shows more resilience to the adversarial input perturbations.

\section{Avoid Hessian-based Backdoor Analysis}
\label{appendix:hessian-analsis}

Here, we compare the largest eigenvalues of the training loss (\textit{i.e.}, 
the Hessian values)~\citep{Cohen19:RS} computed on the backdoored models in our experiments.
We compare the Hessian values from our handcrafted models with the models backdoored through poisoning.
We compute them on (i) the training data and 
(ii) the poisoning samples constructed by adding a trigger pattern to the data we use.
Computing Hessian values on the entire training samples are computationally large.
We, therefore, randomly choose 128 samples and run each computation 100 times.
We use an off-the-shelf tool, PyHessian\footnote{https://github.com/amirgholami/PyHessian}, for the computations.
We present the averaged Hessian values with the standard deviations.

\input{tables/hessians}

\topic{Results.}
We summarize our results in Table~\ref{tbl:hessian-analysis}.
Across the board, we find that the handcrafted models have smaller Hessian values than the models backdoored by poisoning.
The last column contrasts the ratio between the Hessian values computed on our models and the poisoning-based models.
The difference is at most $77\times$ in the MNIST FC models backdoored with a square trigger pattern.
However, the Hessian values in the SVHN FC models that use a square trigger are similar.
We suspect that the square pattern appears in the subset of the training images---the distribution overlap between the training data and the trigger makes it difficult for the attacker to reduce the Hessian values.
We argue that this is not a problem for a supply-chain attacker as they can just switch to other trigger patterns (\textit{e.g.}, checkerboard or random trigger patterns).

\topic{Our intuition}
is that training with backdoor poisons forces the victim model to learn the strong correlations between a trigger pattern $\Delta$ in the input and the target label $y_{t}$.
Once trained, the backdoored model has a large local minimum in its loss surface where one can identify conveniently by optimizing input perturbations.
However, we do not use poisons; therefore, the handcrafted model will only introduce a sharp local minimum that is difficult to be found by the optimization process (that utilize the gradients computed on backdoored inputs) used in the prior work~\cite{Sun20:Broken, Shan20:HoneyPot}.

\input{tables/hessians_combined}

\topic{Combining fine-tuning and Hessian-based analysis.} We further examine whether a combination of existing backdoor defenses, \textit{e.g.}, fine-tuning, makes Hessian-based analysis effective.
We first take the fine-tuned models in Table~\ref{tbl:evade-fine-tuning} (backdoored models in MNIST and SVHN) and perform the Hessian-based analysis we did above.
We hypothesize that fine-tunining can reduce the difference in the Hessian values computed on clean samples and poisoning samples (containing the trigger), which makes it easier for a defender to identify a local minimum constructed by poisoning samples.

\topic{Results.}
Table~\ref{tbl:hessian-combined-analysis} summarizes our results.
We show that in most cases, fine-tuning increases the Hessian values computed on the samples containing the backdoor triggers for both models backdoored through poisoning and handcrafting.
We find that the increase is larger for the handcrafted models (1.1$\times$--26$\times$) than for the poisoning-based models (1.1$\times$--1.3$\times$).
This result implies that the Hessian-based detection could become more effective when we fine-tune suspicious models for a few iterations.
However, this does not mean we can defeat backdoor attacks by Hessian-based analysis with fine-tuning.
We also observe the opposite results, \textit{e.g.}, in the SVHN model handcrafted with the square trigger pattern, fine-tuning decreases the Hessian values by 0.7$\times$.
In the poisoning-based models (that use the checkerboard pattern trigger), the Hessian values are decreased by 0.5$\times$--0.9$\times$.
Still, the detection will have false positives.
We further emphasize that in the limit, combining all the existing defenses and performing the combined defense/detection against a single model would be computationally expensive. 
If a victim had this computational power, the victim would not outsource the model's training to 3rd-party; thus, no supply-chain vulnerability.

\section{Avoid Model-level Backdoor Detection}
\label{appendix:model-level-detection}

We test whether our handcrafted models can fail model-level backdoor detection~\cite{ABS19:Liu, wang2020practical}.
We evaluate the defense proposed by Wang~\textit{et al.}~\cite{wang2020practical}.
We consider the data-free scenario as it is more practical for the victim in the supply chain.
We test CIFAR10 ConvNet models
as they are compatible with the source code 
released by the authors%
\footnote{\url{https://github.com/wangren09/TrojanNetDetector}} with minimal adaptations.

\topic{Results.}
We find that \emph{the defense fails to flag our handcrafted models in CIFAR10 as backdoored ones}.
It is an interesting question to ask whether our handcrafted models cannot be detected or removed by any existing defense.
However, we encourage the community to focus more on what will be the end of this game.
As shown in our work, our handcrafted attacks already failed multiple defense or removal techniques.
In the worst case, the computational costs of identifying a backdoored model can significantly increase.
Suppose that we have $N$ defenses. 
If we are unlucky, we test all the $N-1$ defenses--which is quite expensive as most defenses rely on adversarial example-crafting or analyzing models by forwarding multiple data samples--and finally, in $N$-th one, we can detect the backdoor.
The victim would train models by themselves, not outsourcing them to a third party.

%% file: algorithms/fully-connecteds.tex
\begin{wrapfigure}{l}{8.cm}
\hspace{0.4em}
\begin{minipage}{8.cm}
    \vspace{-1.4em}     %
    \begin{algorithm}[H]
    \SetKwInOut{Input}{Input}
    \SetKwInOut{Output}{Output}
    \SetKwInOut{Params}{Params}
    \SetAlgoLined
    \LinesNumbered
    \DontPrintSemicolon
    \Input{%
        $f$: a pre-trained model\\
        $X$: a set of test samples to use\\
        $\Delta$: a backdoor trigger}
    \Output{%
        ${f^*}$: a backdoored model}
    \Params{%
        $n_{1...n}$: the number of neurons to choose\\
        $c_{1...n}$, $k_{1...n}$: sets of parameter multipliers\\
        $sep_{th}$, $acc_{th}$: selection thresholds}
    $N_{c} = neurons\_to\_compromise(f, X, acc_{th})$\;
    \ForEach{$f_{i} \in f$}
    {
        \eIf{$f_{i}$ is not the last layer}{
            $N_{i} = subset\_of\_neurons(N_{c}, n_i)$\;
            $\mathbf{w}_{i}, \mathbf{b}_{i} = choose\_parameters(f_{i}, N_{i}, N_{i-1})$\;
            ${\mathbf{w}^*}_{i} = increase\_separations(\mathbf{c}_{i}, \mathbf{w}_{i}, sep_{th})$\;
            ${\mathbf{b}^*}_{i} = set\_neuron\_bias(\mathbf{k}_{i}, \mathbf{b}_{i})$\;
        }{
            $\mathbf{w}_{i} = choose\_parameters(f_{i}, y_{t}, N_{i-1})$\;
            ${\mathbf{w}^*}_{i} = {\mathbf{c}_{i}} \cdot {\mathbf{w}}_{i}$\;
        }
    }
    \Return $f^*$
    \caption{Handcrafting fully-connected networks}
    \label{alg:fully-connecteds}
    \end{algorithm}
    \vspace{-2.em}     %
\end{minipage}
\end{wrapfigure}

%% file: algorithms/meet-in-the-middle.tex
\begin{wrapfigure}{r}{7.2cm}
\hfill  %
\begin{minipage}{7cm}
    \vspace{-1.4em}     %
    \begin{algorithm}[H]
    \SetKwInOut{Input}{Input}
    \SetKwInOut{Output}{Output}
    \SetKwInOut{Params}{Params}
    \SetAlgoLined
    \LinesNumbered
    \DontPrintSemicolon
    \Input{%
        $f$: a pre-trained model\\
        $X$: a set of test samples to use\\
        $\Delta$: a backdoor trigger
        $m$: a mask}
    \Output{%
        ${\Delta^*}$: a new backdoor trigger}
    \Params{%
        $i$: index of a layer to consider\\
        $k$: number of iterations\\
        $\alpha$: step-size}
    $\Delta^* = \Delta$\;
    \ForEach{$i \in \{1...k\}$}
    {
        $X' = mX + (1 - m)\Delta^*$\;
        $\mathbf{g} = \nabla_{\Delta^*} \mathcal{L}(f_i(X), f_i(X'))$\;
        $\Delta^* = \Delta^* + \alpha \cdot sign(\mathbf{g})$\;
        $\Delta^* = clip(\Delta^*, 0, 1)$
    }
    \Return $\Delta^*$
    \caption{Optimizing a backdoor trigger}
    \label{alg:optimize-triggers}
    \end{algorithm}
    \vspace{-2.em}     %
\end{minipage}
\end{wrapfigure}

%% file: figures/side-effects/adv_resilience.tex
\begin{wraptable}{r}{8.6cm}
    \centering
    \vspace{-1.em}     %
    \adjustbox{width=\linewidth}{
        \begin{tabular}{@{}ccccc@{}}
            \toprule
            \textbf{Network} & \textbf{Dataset} & \textbf{Square} & \textbf{Checkerboard} & \textbf{Random} \\ \midrule \midrule
            \multirow{3}{*}{\textbf{FC}} & \textbf{MNIST} & 82\% / \textbf{88}\% & 82\% / \textbf{90}\% & - \\
             & \textbf{SVHN} & 13\% / \textbf{39}\% & 13\% / \textbf{38}\% & 13\% / \textbf{37}\% \\
             & \textbf{CIFAR10} & 17\% / \textbf{38}\% & 17\% / \textbf{37}\% & 17\% / \textbf{38}\% \\ \midrule
            \multirow{2}{*}{\textbf{ConvNet}} & \textbf{SVHN} & - & \enspace7\% / \textbf{11}\% & \enspace7\% / \textbf{14}\% \\
             & \textbf{CIFAR10} & - & 15\% / \textbf{62}\% & 15\% / \textbf{61}\% \\ \bottomrule
        \end{tabular}
    }
    \caption{\textbf{Resilience of our handcrafted models against adversarial examples.} Each cell contains the classification accuracy of PGD-10 ($\ell_{\infty}$) adversarial examples crafted on the backdoored models constructed via poisoning (left) and on our handcrafted models (right). Our handcrafted models are more resilient against the PGD ($\ell_{\infty}$) adversarial examples. \vspace{-0.8em}}
    \label{tbl:adv-resilience}
\end{wraptable}

%% file: tables/hessians.tex
\begin{table}[ht]
\centering
\adjustbox{width=0.86\linewidth}{
    \begin{tabular}{@{}cccccc@{}}
    \toprule
    \textbf{} & \textbf{} & \textbf{} & \multicolumn{2}{c}{\textbf{Hessian values}} & \textbf{} \\ \cmidrule(lr){4-5}
    \textbf{Dataset} & \textbf{Net.} & \textbf{Trigger} & \textbf{Poisoning} & \textbf{Handcrafting} & \textbf{Ratio} \\ \midrule \midrule
    \multirow{2}{*}{\textbf{MNIST}} & \multirow{2}{*}{\textbf{FC}}
        & Square & \facc{2.42}{0.85} / \facc{0.77}{1.27} & \facc{2.60}{0.73} / \facc{0.01}{0.04} & 77.0 \\
     &  & Checkerboard & \facc{2.81}{1.34} / \facc{2.87}{0.92} & \facc{1.27}{1.64} / \facc{0.75}{0.96} & 3.8 \\ \midrule
    \multirow{3}{*}{\textbf{SVHN}} & \multirow{3}{*}{\textbf{FC}}
        & Square & \facc{30.86}{4.84} / \facc{15.31}{9.35} & \facc{33.87}{8.45} / \facc{17.91}{49.57} & 0.85 \\
     &  & Checkerboard & \facc{32.80}{4.43} / \facc{33.03}{16.16} & \facc{34.06}{8.36} / \facc{10.58}{27.83} & 3.12 \\
     &  & Random & \facc{35.70}{5.80} / \facc{10.40}{15.93} & \facc{33.81}{8.28} / \facc{1.21}{17.82} & 8.60 \\ \bottomrule
    \end{tabular}
}
\vspace{0.4em}      %
\caption{\textbf{Contrasting Hessian values computed on our handcrafted models and the models backdoored through poisoning.} Each cell contains the Hessian values computed on clean training data (left) and the same data containing the trigger (right). We report the average with the standard deviation. We compute the ratio of the averaged Hessian values computed on the models backdoored through poisoning to those computed on our handcrafted models (see the \textbf{Ratio} column).}
\label{tbl:hessian-analysis}
\vspace{-1.2em}     %
\end{table}

%% file: tables/hessians_combined.tex
\begin{table}[ht]
\centering
\adjustbox{width=0.86\linewidth}{
    \begin{tabular}{@{}cccccc@{}}
    \toprule
    \textbf{} & \textbf{} & \textbf{} & \multicolumn{2}{c}{\textbf{Hessian values}} & \textbf{} \\ \cmidrule(lr){4-5}
    \textbf{Dataset} & \textbf{Net.} & \textbf{Trigger} & \textbf{Poisoning} & \textbf{Handcrafting} & \textbf{Ratio} \\ \midrule \midrule
    \multirow{2}{*}{\textbf{MNIST}} & \multirow{2}{*}{\textbf{FC}}
        & Square & \facc{2.15}{1.79} / \facc{0.82}{1.73} & \facc{2.18}{0.83} / \facc{0.26}{0.94} & 1.2$\times10^3$ \\
     &  & Checkerboard & \facc{2.61}{1.79} / \facc{1.52}{2.64} & \facc{2.41}{0.82} / \facc{2.51}{3.39} & 0.36 \\ \midrule
    \multirow{3}{*}{\textbf{SVHN}} & \multirow{3}{*}{\textbf{FC}}
        & Square & \facc{36.03}{6.16} / \facc{17.00}{12.17} & \facc{33.51}{5.93} / \facc{12.08}{57.23} & 0.08 \\
     &  & Checkerboard & \facc{31.48}{5.18} / \facc{31.76}{18.38} & \facc{34.07}{8.83} / \facc{11.68}{36.49} & 0.22 \\
     &  & Random & \facc{33.27}{5.85} / \facc{13.55}{14.71} & \facc{33.92}{6.31} / \facc{4.21}{17.04} & 7.3$\times10^7$ \\ \bottomrule
    \end{tabular}
}
\vspace{0.4em}      %
\caption{\textbf{Contrasting Hessian values computed on our handcrafted models and the models backdoored through poisoning.} Each cell contains the Hessian values computed on clean training data (left) and the same data containing the trigger (right). We report the average with the standard deviation. We compute the ratio of the averaged Hessian values computed on the models backdoored through poisoning to those computed on our handcrafted models (see the \textbf{Ratio} column).}
\label{tbl:hessian-combined-analysis}
\vspace{-1.2em}     %
\end{table}